\documentclass[aps,prb,reprint,superscriptaddress, floatfix,longbibliography,nofootinbib]{revtex4-2}

\usepackage{amsmath,amssymb,amsfonts}
\usepackage{bm}
\usepackage{graphicx}
\usepackage{comment}
\usepackage{dsfont}
\usepackage{xcolor}
\usepackage[
  colorlinks=true,
  linkcolor=blue,
  citecolor=blue,
  urlcolor=blue
]{hyperref}

\newcommand{\Tr}{\operatorname{Tr}}
\newcommand{\diag}{\operatorname{diag}}
\newcommand{\half}{\tfrac{1}{2}}
\newcommand{\LL}{Luttinger-liquid}

\newcommand{\XY}{\mathrm{XY8}}
\renewcommand{\Re}{\mathrm{Re}\,}
\renewcommand{\Im}{\mathrm{Im}\,}

\begin{document}

\title{Probing Spatiotemporal Quantum Correlations with Two-Qubit Relaxometry}

\author{Christina Psaroudaki}
\affiliation{Laboratoire de Physique de l'\'Ecole Normale Sup\'erieure, ENS,
Universit\'e PSL, CNRS, Sorbonne Universit\'e, Universit\'e de Paris,
F-75005 Paris, France}

\date{\today}

\begin{abstract}
Quantum impurities and spin qubits are powerful local probes of magnetic fluctuations, but conventional relaxometry reduces their information to a local relaxation rate at a single frequency. Building on advances in two-qubit sensing of spatial correlations, we show how two complementary protocols probe finite-time correlations in interacting quantum matter. Phase-cycled Bell-state relaxometry isolates the signed contribution of correlated relaxation to the joint ground-state population. To second order in the probe--environment coupling, the resulting population contrast measures a finite-time integral of the energy-ordered cross-correlator. In the longitudinal regime, XY8 covariance probes the complementary symmetrized cross-correlator within a tunable frequency window. For a gapless quantum spin chain, the signals reveal interaction-dependent algebraic scaling and ballistic propagation. Luttinger-liquid theory links sign changes as the qubit frequency is swept to the propagation time between the probes. It also predicts distinct spatial power laws for the Bell signal before the correlation front and at late times, when the sensing window resolves the qubit frequency. Matrix-product-state calculations for the spin chain test the pre-front scaling at zero and finite magnetization and extract the propagation velocity from Bell oscillations at zero magnetization. The probe coupling geometry determines which spin and momentum sectors contribute to the measured signals. \end{abstract}

\maketitle

\section{Introduction}
\label{sec:intro}

Strongly correlated quantum systems are governed by the dynamics of their many-body correlations \cite{DeChiara_2018}. Although theory can predict universal scaling and characteristic propagation velocities, direct experimental access to real-space, finite-time quantum correlations remains limited \cite{PhysRevB.92.180205,PhysRevB.96.014204,PhysRevLett.111.147205}. Resolving how these correlations are distributed and propagate in space and time is therefore a central challenge in quantum many-body physics.

Quantum impurities and spin qubits provide local probes of magnetic fluctuations with high spatial resolution, tunable frequency selectivity, and weak back-action. In relaxometry, the probe transition rate is set by the dynamical spin correlations of the surrounding quantum environment, making the qubit a frequency- and momentum-selective detector of magnetic noise \cite{PhysRevLett.121.187204,PhysRevB.98.195433,PhysRevB.99.104425,Casola2018}. Impurity atoms immersed in ultracold gases \cite{Adam2022,Streif2016} and solid-state defect spins operated as effective spin-$1/2$ sensors \cite{RevModPhys.92.015004,Hall2016,Casola2018} are established realizations. Qubit dephasing spectroscopy has further been proposed to extract critical exponents, conservation laws, and dynamical universality classes \cite{PhysRevLett.131.070801,gmpy-2sgh}, while single-qubit decoherence can encode phase boundaries, ground-state entanglement, and propagation velocities in strongly interacting spin systems \cite{PhysRevB.111.L161115,Das2026Thermal}. Quantum sensors thus provide microscopic access to interacting matter, but predominantly of fluctuations sampled locally at a single probe position.

Beyond this spatial limitation, conventional relaxometry usually invokes the Fermi-golden-rule limit, in which long interrogation times collapse the finite-bandwidth response onto the bath spectrum at the qubit transition frequency. This Markovian description discards the finite-time structure of the underlying correlations, which can remain important in interacting quantum environments \cite{PhysRevB.111.L161115,Das2026Thermal}. Such information is particularly relevant for spatially separated probes, where finite interrogation times can resolve the propagation of correlations between the two sensing regions. While finite-time filter-function approaches are well established for dephasing spectroscopy \cite{PhysRevB.77.174509,PhysRevLett.131.070801,PhysRevB.111.L161115,gmpy-2sgh,PhysRevA.109.032412}, their relaxation-sector counterpart has received less attention, and finite-time kernels become indispensable when the propagation time between spatially separated probes is itself part of the signal.

Multi-qubit probes have already been used to access spatially correlated noise through dynamical-decoupling spectroscopy \cite{Paz-Silva_2016,PhysRevA.94.012109,Krzywda_2019,Zou2024,Gulacsi2025}, superconducting-qubit cross-noise measurements \cite{PRXQuantum.1.010305}, and two-NV covariance magnetometry \cite{Rovny2022}. Related studies have examined the effects of spatially correlated noise on driven-qubit dynamics \cite{Zou2024} and proposed collective relaxation as a probe of noise correlations \cite{Gulacsi2025}. These approaches characterize environmental fluctuations through prescribed classical or quantum noise cross-spectra.

These advances establish two-qubit sensing as a probe of spatially correlated environments, primarily through longitudinal coupling and often within a Gaussian description. A remaining challenge is to connect the measured signals directly to the microscopic correlations of strongly interacting quantum matter, whose local observables are non-Gaussian and noncommuting and whose correlations exhibit intrinsic critical scaling and propagation. Here, we address this challenge by retaining the full physical qubit--environment coupling and deriving the cross spectra governing both relaxation and dephasing microscopically from the nonlocal dynamical spin correlations of the chain.

At finite sensing times, the probe dynamics separates into local decoherence and cross-kernels that encode correlations between the two sensing regions. To second order in the probe coupling, both contributions follow from environmental two-point functions without a Markov or Gaussian approximation. The two protocols probe different cross-kernels. Cycling the relative phase of a one-excitation Bell state isolates the signed contribution of correlated relaxation to the joint ground-state population, giving access to the energy-ordered cross-correlator. In the longitudinal regime, XY8 covariance probes the complementary symmetrized cross-correlator within a tunable frequency window.

We apply these protocols to the gapless spin-$1/2$ chain, where Luttinger-liquid theory relates the algebraic spatial decay of the signals to the Luttinger parameter. In Bell relaxometry, the spacing of sign changes as the qubit frequency is swept reflects the propagation time between the probes. Matrix-product-state calculations test the spatial scaling of the Bell signal before the correlation front at zero and finite magnetization. At zero magnetization, we extract the propagation velocity from oscillations of the Bell cross-relaxation rate after the front. The coupling geometry determines which spin and momentum sectors contribute to the signals of lattice-scale impurity qubits and defect spins.

Section~\ref{sec:model} develops the finite-time reduced dynamics of two independently controlled probes, and Sec.~\ref{sec:protocols} introduces Bell-state relaxometry and XY8 covariance. We specialize to the interacting spin chain in Sec.~\ref{sec:spinchain} and analyze the spatial scaling and propagation signatures of the two protocols in Secs.~\ref{sec:Bell_LL} and \ref{sec:XY8_LL}. Section~\ref{sec:exp} discusses implementations, and Sec.~\ref{sec:conclusions} concludes.
Appendix~\ref{app:single} presents the corresponding single-probe theory, and the remaining appendices give technical details.

\begin{figure}[t]
\centering
\includegraphics[width=\linewidth]{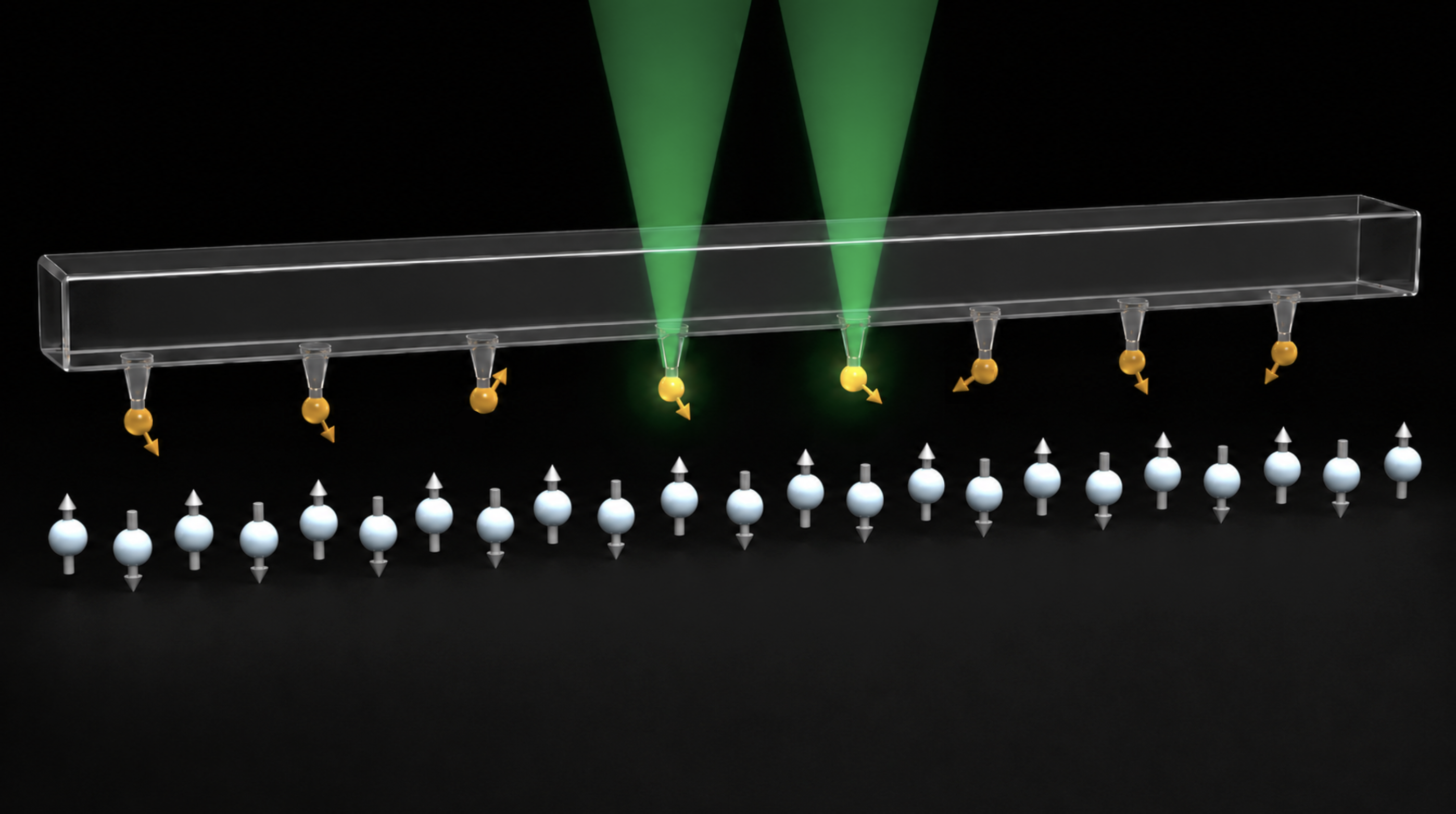}
\caption{\label{fig:schematic}%
Two-probe relaxometry of an interacting spin chain. An array of probe qubits (orange) couples to a quantum spin chain (blue) with spatial range $h$. Two qubits separated by $d$ are independently controlled, and their connected signal probes nonlocal dynamical spin correlations between the sensing regions.}
\label{fig:Couplig_Scheme}
\end{figure}

\section{Two-probe sensing of quantum matter}
\label{sec:model}

We consider two probe qubits, denoted $A$ and $B$, weakly coupled to the same interacting quantum system and sampling two spatially separated regions. The essential feature relative to single-probe relaxometry is that the reduced two-qubit dynamics contains, in addition to local self-kernels, genuine cross-kernels controlled by nonlocal quantum correlation functions between the regions sampled by the two probes \cite{PhysRevB.92.180205,PhysRevA.94.012109}. These cross-kernels are the objects from which spatial and spatiotemporal correlations of the quantum matter can be extracted.

Controlled multi-qubit dynamics under spatially correlated dephasing has been treated within filter-function approaches \cite{Paz-Silva_2016,Krzywda_2019}, and two-qubit Ramsey and spin-echo spectroscopy of a many-body environment, with longitudinal coupling and a Gaussian weak-coupling evaluation, in Ref.~\cite{Hosseinabadi2026}. At the same time, second-order descriptions of driven qubits with correlated quantum noise have been developed recently \cite{Zou2024,Gulacsi2025}. Two-qubit spectroscopy of spatially correlated noise has been demonstrated experimentally \cite{PRXQuantum.1.010305}. These approaches characterize the environment through prescribed cross-spectra, often for single-axis coupling. Here, we formulate two-probe sensing directly in terms of the microscopic correlations of interacting quantum matter, with general tensorial probe--matter coupling and without assuming Gaussian bath statistics. The resulting cross-kernels connect the measured two-qubit signals to the intrinsic critical scaling and light-cone structure of the many-body system.

\subsection{General model}
\label{sec:model_def}

We consider two independently controlled spin-$1/2$ probes, $A$ and $B$, coupled to spatially separated regions of a common quantum many-body system, shown in Fig.~\ref{fig:Couplig_Scheme}. The probe--matter coupling is weak, while the interactions within the many-body system are unrestricted. The total Hamiltonian is 
\begin{equation}
H(t)=H_0(t)+V,\qquad
H_0(t)=H_M+H_Q^{(A)}(t)+H_Q^{(B)}(t),
\label{eq:Htot}
\end{equation}
where $H_M$ governs the quantum matter, and we use $\hbar=1$ throughout. Each probe has an independently
controlled Hamiltonian,
\begin{equation}
H_Q^{(a)}(t)=H_{Q,0}^{(a)}(t)+H_c^{(a)}(t),
\qquad a\in\{A,B\},
\label{eq:HQa}
\end{equation}
with $H_{Q,0}^{(a)}$ the free qubit Hamiltonian and $H_c^{(a)}$ the applied control Hamiltonian. We assume no direct coupling between the probes, so that any interaction generated between them is mediated by the quantum matter.
\begin{figure*}[t]
\centering
\includegraphics[width=0.9\linewidth]{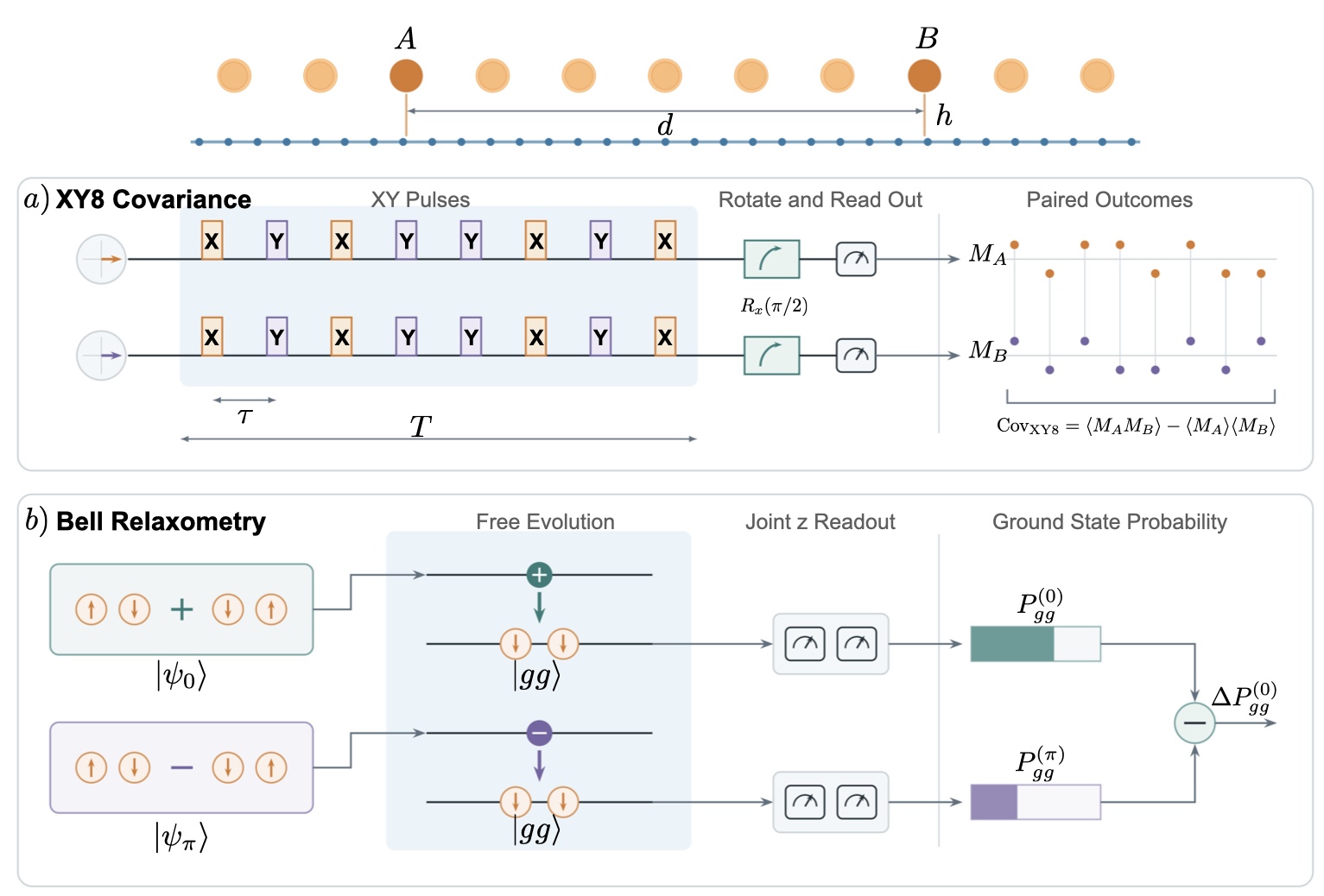}
\caption{\label{fig:exp_setup}\textbf{Two-probe sensing of correlations in a quantum spin chain.}
Two probe qubits, $A$ and $B$ (filled orange circles), are separated by $d$ and couple to the spin chain (blue). (a) \textit{XY8 covariance}. Similar to \cite{Rovny2022}, both probes start in $\lvert +x\rangle$. During the sensing time $T$, each receives $\pi$ pulses about $x$ and $y$, spaced by $\tau$. A final $\pi/2$ rotation about $x$ maps $S^y$ onto the $z$-basis readout. The connected covariance of each pair of  $M_A,M_B=\pm 1/2$ from every repetition probes the symmetrized cross-correlations between the two sensing regions near $\omega_\star\simeq\pi/\tau$. (b) \textit{Bell-state relaxometry}. Separate runs start in the symmetric $|\psi_0\rangle$ or antisymmetric $|\psi_\pi\rangle$ state. Either probe can transfer its excitation to the chain, leaving the pair in $\lvert gg\rangle$. Joint readout after time $T$ gives the fraction $P_{gg}^{(\phi)}$ of such outcomes. The half-difference $\Delta P_{gg}^{(0)}=[P_{gg}^{(0)}-P_{gg}^{(\pi)}]/2$ removes the phase-independent background and probes the unsymmetrized cross-correlations near the probe splitting $\omega_q$. The bars illustrate a positive signal, for which the symmetric state decays faster.}
\end{figure*}

We write the probe--matter interaction as
\begin{equation}
V=V_A+V_B,\qquad V_a=\sum_{\mu=x,y,z} S_a^\mu\otimes\delta\mathcal O_a^\mu,
\label{eq:Va_def}
\end{equation}
where $S_a^\mu=\sigma_a^\mu/2$ acts on probe $a$ and
\begin{equation}
\delta\mathcal O_a^\mu
=\mathcal O_a^\mu-\langle\mathcal O_a^\mu\rangle_M
\label{eq:O_connected}
\end{equation}
is the fluctuation of the matter operator coupled to that spin component. The operators $\mathcal O_a^\mu$ are arbitrary Hermitian operators that incorporate the tensorial structure and spatial profile of the coupling. For coupling to local lattice operators $O_j^\lambda$, one may write,
\begin{equation}
\mathcal O_a^\mu
=\sum_{j,\lambda}g_{a,j}^{\mu\lambda}O_j^\lambda,
\label{eq:generic_spatial_coupling}
\end{equation}
where $j$ labels lattice sites, $\lambda$ labels matter-operator
components, and $g_{a,j}^{\mu\lambda}$ specifies the coupling tensor
and its spatial dependence. The mean-field term
$\sum_\mu\langle\mathcal O_a^\mu\rangle_M S_a^\mu$
is absorbed into $H_{Q,0}^{(a)}$, leaving only the connected
fluctuations in Eq.~\eqref{eq:Va_def}. We assume a factorized initial state,
\begin{equation}
\rho_{\mathrm{tot}}(0)=\rho_{AB}(0)\otimes\rho_M, \qquad [\rho_M,H_M]=0,
\label{eq:init}
\end{equation}
where $\rho_{AB}(0)$ is an arbitrary two-qubit state and $\rho_M$ is a stationary state of the quantum matter.
\subsection{Control-dressed kernels}
\label{sec:kernels}

Because the controls act independently on the two probes and there is no direct bare coupling between them, the uncoupled qubit propagator factorizes exactly,
\begin{equation}
U_Q(t,0)=U_A(t,0)\otimes U_B(t,0),
\label{eq:UQ_factorize}
\end{equation}
with
$U_a(t,0)=\mathcal T\exp\big[-i\int_0^t ds\,H_Q^{(a)}(s)\big]$. Passing to the interaction picture with respect to $H_0(t)$ gives
\begin{equation}
\widetilde V(t)=\sum_{a=A,B}\sum_{\mu=x,y,z}\widetilde S_a^\mu(t)\otimes\delta\widetilde{\mathcal O}_a^\mu(t),
\end{equation}
where $\widetilde S_a^\mu(t)=U_a^\dagger(t,0)S_a^\mu U_a(t,0)$ and $\delta\widetilde{\mathcal O}_a^\mu(t) =e^{iH_Mt}\delta\mathcal O_a^\mu e^{-iH_Mt}$. Since each probe is a spin-$1/2$, the adjoint action of $U_a$ is an exact $SO(3)$ rotation of the spin vector \cite{Green_2013},
\begin{equation}
\widetilde S_a^\mu(t) = \sum_{\nu=x,y,z}R_{\mu\nu}^{(a)}(t)\,S_a^\nu, \qquad R^{(a)}(t)\in SO(3). \label{eq:Ra_def}
\end{equation}
All control dependence, including pulse sequences, is therefore encoded in the two independent matrices $R^{(A)}(t)$ and $R^{(B)}(t)$. Their factorization into free and pulse segments, and the ideal-pulse limit, are given in Appendix~\ref{app:single}, Sec.~\ref{app:pulse}.

The many-body correlations sampled by the probes are
\begin{equation}
C_{ab}^{\mu\nu}(t_1,t_2) = \Big\langle \delta\widetilde{\mathcal O}_a^\mu(t_1) \delta\widetilde{\mathcal O}_b^\nu(t_2) \Big\rangle_M, \qquad a,b\in\{A,B\}.
\label{eq:Cab_def}
\end{equation}
The diagonal correlators $C_{AA}$ and $C_{BB}$ describe fluctuations sampled by each probe individually, whereas $C_{AB}$ and $C_{BA}$ contain correlations between the two sensing regions. If the operators $\mathcal O_A^\mu$ and $\mathcal O_B^\mu$ are centered around $\mathbf R_A$ and $\mathbf R_B$, respectively, the cross-correlators depend on the probe separation $d=|\mathbf R_A-\mathbf R_B|$. 

The corresponding control-dressed kernels are
\begin{equation}
\mathcal K_{ab}^{\mu\nu}(t_1,t_2) = \sum_{\mu',\nu'} R_{\mu'\mu}^{(a)}(t_1) R_{\nu'\nu}^{(b)}(t_2) C_{ab}^{\mu'\nu'}(t_1,t_2)\,,
\label{eq:KAB_def}
\end{equation}
where the many-body dynamics enters through the correlation functions $C_{ab}^{\mu\nu}$, the tensorial and spatial structure of the coupling is contained in $\mathcal O_a^\mu$, and the applied control enters through $R^{(a)}$. 
\subsection{Finite-time time-convolutionless dynamics}
\label{sec:tcl_form}
We treat the probe--matter interaction perturbatively using the time-convolutionless projection operator formalism \cite{Chaturvedi1979,PhysRevLett.93.130406,Breuer2016,Zou2024}. Here we retain a general tensorial coupling to an interacting quantum many-body system and keep the resulting microscopic cross-correlators explicitly, while details of the derivation can be found in the Appendix ~\ref{app:dyson}. Expanding the interaction-picture reduced map to second order in $V$,
\begin{equation}
\widetilde\rho_{AB}(t)= \Phi_{AB}(t)\rho_{AB}(0) =
\big[1+\Phi_2^{AB}(t)\big]\rho_{AB}(0),
\end{equation}
and defining the time-convolutionless generator
$\mathcal G_{AB}=\dot\Phi_{AB}\Phi_{AB}^{-1}$ \cite{Chaturvedi1979,Breuer2016}, one obtains at second order,
\begin{align}
\mathcal G_{AB}^{(2)}(t)[\rho] =-\sum_{a,b}\sum_{\mu,\nu}\int_0^t dt'\Big[
&\mathcal K_{ab}^{\mu\nu}(t,t')[S_a^\mu,S_b^\nu\rho]\nonumber\\
+&\mathcal K_{ba}^{\nu\mu}(t',t)[\rho S_b^\nu,S_a^\mu]\Big].
\label{eq:TCL2_gen_AB}
\end{align}
The reduced dynamics therefore obeys
\begin{equation}
\dot{\widetilde\rho}_{AB}(t)=\mathcal G_{AB}^{(2)}(t)\widetilde\rho_{AB}(t)
+O(V^3).
\label{eq:TCL2_AB}
\end{equation}

Equation~\eqref{eq:TCL2_AB} is time-local in the reduced state but does not invoke a Markov approximation, and its generator retains finite-time bath correlations through the integral over $0\leq t'\leq t$. The importance of this temporal structure has already been demonstrated for a spin qubit locally coupled to an XXZ chain, whose dephasing exhibits non-Markovian dynamics and coherence revivals \cite{PhysRevB.111.L161115,Das2026Thermal}. For the XXZ chain considered below, second-order expansions reproduce the qubit dephasing dynamics from full tensor-network simulations in the gapless regime at weak probe coupling, while higher-order processes become relevant in the gapped phases of the model~\cite{PhysRevB.111.L161115}. Recovering the Markovian Fermi-golden-rule limit requires the additional approximation of replacing the finite-time coefficients by time-independent rates \cite{Breuer2016,PhysRevLett.131.070801,PhysRevB.98.195433}, by extending the memory integral to infinity (see Appendix~\ref{app:fgr} for details).

The structure of Eq.~\eqref{eq:TCL2_gen_AB} separates naturally into probe-space blocks,
\begin{equation}
\mathcal G_{AB}^{(2)}= \mathcal G^{AA}+\mathcal G^{BB}+\mathcal G^{AB}+\mathcal G^{BA}.
\label{eq:blocks}
\end{equation}
The diagonal blocks describe the local dynamics of the individual probes, whereas the off-diagonal blocks are matter-mediated cross-generators governed by $C_{AB}^{\alpha\beta}$ and $C_{BA}^{\alpha\beta}$. It is useful to separate quantum fluctuations from response \cite{RevModPhys.82.1155}. Introducing
\begin{align}
N_{ab}^{\mu\nu}(t_1,t_2)
&=\half \Big\langle \{\delta\widetilde{\mathcal O}_a^\mu(t_1), \delta\widetilde{\mathcal O}_b^\nu(t_2) \} \Big\rangle_M ,
\label{eq:Nab_def} \\
X_{ab}^{\mu\nu}(t_1,t_2) &=\frac{1}{2i} \Big\langle [ \delta\widetilde{\mathcal O}_a^\mu(t_1),
\delta\widetilde{\mathcal O}_b^\nu(t_2)]
\Big\rangle_M ,
\label{eq:Xab_def}
\end{align}
and dressing them with $R^{(a)}$ and $R^{(b)}$ as in Eq.~\eqref{eq:KAB_def}, the generator may equivalently be written
\begin{align}
\mathcal G_{AB}^{(2)}(t)[\rho]=-\sum_{a,b}\sum_{\mu,\nu} \int_0^t dt'
\Big[& \mathcal N_{ab}^{\mu\nu}(t,t')[S_a^\mu,[S_b^\nu,\rho]]\nonumber\\
+&i\mathcal X_{ab}^{\mu\nu}(t,t')[S_a^\mu,\{S_b^\nu,\rho\}]\Big].
\label{eq:TCL2_noise_response}
\end{align}
The first term contains the symmetrized quantum fluctuations, while the second contains the bath response. For later use, the second-order time-convolutionless (TCL2) solution can also be written in exponential form. Since
$\mathcal G_{AB}^{(2)}=\dot\Phi_2^{AB}$ and $\Phi_2^{AB}(0)=0$,
\begin{equation}
\int_0^T ds\,\mathcal G_{AB}^{(2)}(s)=\Phi_2^{AB}(T).
\end{equation}
The first time-ordering correction in the Magnus expansion is $O(V^4)$ and therefore lies beyond our second-order approximation. We may thus replace the time-ordered exponential by an ordinary exponential and write
\begin{equation}
\widetilde\rho_{AB}(T) = e^{\Phi_2^{AB}(T)} \rho_{AB}(0) + O(V^3).
\label{eq:exp_form}
\end{equation}
The exponent is a Liouville-space superoperator built from the control-dressed quantum correlation functions and reduces to a scalar attenuation exponent only in commuting sectors, such as pure dephasing. Exponentiation does not change the perturbative accuracy. 

\section{Two-probe protocols for nonlocal quantum correlations}
\label{sec:protocols}

Spatial correlations in a quantum environment affect both the probes' exchange of energy with the environment and their phases. We use two complementary protocols to resolve these effects, summarized in Fig.~\ref{fig:exp_setup}. Bell-state relaxometry follows the decay of an entangled two-qubit state to isolate the signed contribution of nonlocal correlations to relaxation. It samples the energy-exchange component of the full cross-correlator $N_{AB}+iX_{AB}$. Dynamical-decoupling covariance compares the readouts of two independently controlled probes to reveal shared phase fluctuations and samples the symmetrized cross-correlator $N_{AB}$ of Eq.~\eqref{eq:Nab_def} within a tunable frequency window. The cross-response $X_{AB}$ can be accessed separately by polarizing one probe and reading the induced phase of the other \cite{Hosseinabadi2026}, a preparation that breaks the probe-exchange symmetry used below. We derive both signals for general quantum matter, retaining finite sensing times and accumulated local decoherence, before applying them to an interacting spin chain in Sec.~\ref{sec:spinchain}.
\subsection{Bell-state relaxometry}
\label{sec:Bell}

Bell-state relaxometry isolates the interference between the two pathways by which a probe transfers its excitation to the environment\cite{PhysRev.93.99}. For identical, freely precessing probes, the half-difference of the joint ground-state populations measured at preparation phases $\phi$ and $\phi+\pi$ is
\begin{equation}
\Delta P_{gg}^{(\phi)}(T)= \Re\!\left[e^{-i\phi}W_{BA}^{\downarrow}(T)\right]+O(V^3).
\label{eq:Bell_contrast_second_order}
\end{equation}
The finite-time kernel $W_{BA}^{\downarrow}$ in Eq.~\eqref{eq:Wab_double} samples the energy-ordered cross-correlator for energy transfer to the environment near $\omega_q$. For exchange-equivalent probes, $W_{BA}^{\downarrow}=W_{AB}^{\downarrow}\in\mathbb R$, so the $\phi=0$ contrast gives the signed cross-kernel to second order.

For the interrogation interval we take identical, freely precessing probes,
\begin{equation}
H_Q^{(a)}= \omega_q S_a^z,
\end{equation}
so that the uncoupled propagator $U_Q(T,0)$ of Eq.~\eqref{eq:UQ_factorize}
reduces to $e^{-i\omega_qT(S^z_A+S^z_B)}$ and commutes with the projector $\Pi_{gg}=|gg\rangle\langle gg|$, which may therefore be used equally in the Schr\"odinger and interaction pictures. Under a general control protocol it would be replaced by $\widetilde\Pi_{gg}(T)=U_Q^\dagger(T,0)\,\Pi_{gg}\,U_Q(T,0)$.

We denote by $|g\rangle$ and $|e\rangle$ the eigenstates of $S^z_a$ with
eigenvalues $\mp 1/2$, and write two-qubit product states as $|ij\rangle=|i\rangle_A\otimes|j\rangle_B$. In the one-excitation subspace spanned by $|eg\rangle$ and $|ge\rangle$ we prepare the Bell state
\begin{equation}
|\psi_\phi\rangle=\frac{1}{\sqrt2} \left(|eg\rangle+e^{i\phi}|ge\rangle\right),
\qquad
\rho_\phi(0)=|\psi_\phi\rangle \langle\psi_\phi|,
\label{eq:bell_state}
\end{equation}
and measure the probability of finding both probes in the ground state,
\begin{equation}
P_{gg}^{(\phi)}(T)= \Tr [\Pi_{gg}e^{\Phi_2^{AB}(T)}\rho_\phi] =P_0(T)+
\Re[e^{-i\phi}\mathcal C_{AB}(T)],
\label{eq:Pgg}
\end{equation}
with
\begin{align}
P_0(T)=\half\Tr\Big[\Pi_{gg}e^{\Phi_2^{AB}}\big(|eg\rangle\langle eg|
+|ge\rangle\langle ge|\big)\Big]
\end{align}
and
\begin{equation}
\mathcal C_{AB}(T)=\Tr\big[\Pi_{gg}e^{\Phi_2^{AB}(T)}|eg\rangle\langle ge|\big].
\label{eq:CAB}
\end{equation}
Here, the quantity $\mathcal C_{AB}(T)$ is the phase-sensitive, matter-mediated contribution to the signal. It contains the full TCL2 evolution generated jointly by dephasing, relaxation, and the nonlocal $AB$ and $BA$ kernels.

The phase-independent background is eliminated without any expansion by
combining two Bell phases differing by $\pi$,
\begin{equation}
\Delta P_{gg}^{(\phi)}(T)\equiv\half\left[P_{gg}^{(\phi)}-P_{gg}^{(\phi+\pi)}\right]
=\Re\big[e^{-i\phi}\mathcal C_{AB}(T)\big].
\label{eq:Bell_phase_difference_general}
\end{equation}
Thus, $\Delta P_{gg}^{(0)}=\Re\,\mathcal C_{AB}$ and $\Delta P_{gg}^{(\pi/2)}=
\Im\,\mathcal C_{AB}$. Correlated relaxation can already be detected from population dynamics through collective emission \cite{Gulacsi2025}, complex two-qubit noise cross spectra have been reconstructed experimentally using spin-locking spectroscopy and state tomography \cite{PRXQuantum.1.010305}, and parity-resolved
relaxation rates of a double quantum dot can recover signed dynamical
spin correlations \cite{MOHAMED2027419514}. Here the relative Bell phase isolates the matter-mediated cross signal from the phase-independent local background.

Suppose a symmetry of the dynamics exchanges the probes, maps $\lvert\psi_\phi\rangle$ to $\lvert\psi_{-\phi}\rangle$ up to an overall phase, and leaves the joint ground-state projector
$\Pi_{gg}$ invariant. Then $P_{gg}^{(\phi)}(T)=P_{gg}^{(-\phi)}(T)$ and $\operatorname{Im}\mathcal{C}_{AB}(T)=0$. The contrast $\Delta P_{gg}^{(0)}(T)=\operatorname{Re}\mathcal{C}_{AB}(T)$ is the half-difference of the joint ground-state populations for the symmetric ($\phi=0$) and antisymmetric ($\phi=\pi$) preparations. The $\pi/2$ quadrature $\Delta P_{gg}^{(\pi/2)}(T)$ vanishes and serves as a null test of this symmetry.

The transverse part of the coupling~\eqref{eq:Va_def} is
\begin{equation}
V_\perp=\frac12\sum_{a=A,B}
\left(
S_a^+\otimes\delta\mathcal O_a^-
+
S_a^-\otimes\delta\mathcal O_a^+
\right),
\end{equation}
where $S_a^\pm=S_a^x\pm iS_a^y$ and
$\delta\mathcal O_a^\pm=\delta\mathcal O_a^x\pm i\delta\mathcal O_a^y$. The operator $S_a^-$ lowers probe $a$ from its excited state to its ground state. Starting from $|\psi_\phi\rangle$, relaxation of either probe therefore leads to $|gg\rangle$. The measured population depends on the individual decay processes and on the interference between these two alternatives.

For freely precessing probes, $S_a^\pm(t)=S_a^\pm e^{\pm i\omega_qt}$. The corresponding finite-time relaxation kernels, obtained from
Eq.~\eqref{eq:TCL2_gen_AB}, are
\begin{align}
W_{ab}^{\downarrow}(T)={}&\frac14\int_0^T dt_1\int_0^T dt_2\,e^{-i\omega_q(t_1-t_2)}\nonumber\\
&\times\left\langle\delta\widetilde{\mathcal O}_a^-(t_2)\delta\widetilde{\mathcal O}_b^+(t_1)\right\rangle_M .
\label{eq:Wab_double}
\end{align}
The kernels $W_{AA}^{\downarrow}$ and $W_{BB}^{\downarrow}$ describe the local relaxation of probes $A$ and $B$, respectively. The cross-kernels $W_{AB}^{\downarrow}$ and $W_{BA}^{\downarrow}$ determine the interference between the two decay pathways and are governed by the nonlocal correlations of the environment. The oscillating factor selects energy transfer near the probe splitting $\omega_q$, with a frequency resolution set by the sensing time $T$.

For a stationary environment, the correlator depends only on the
time difference, and the double integral reduces to
\begin{equation}
W_{ab}^{\downarrow}(T;d_{ab})=\frac14 \int_{-T}^{T} dt(T-|t|)e^{-i\omega_qt}\,
C_{ab}^{\downarrow}(t;d_{ab}).
\label{eq:WAB_finite_time}
\end{equation}
with $C_{ab}^{\downarrow}(\tau)=
\langle\delta\mathcal O_a^-(0)\delta\widetilde{\mathcal O}_b^+(\tau)\rangle_M$. Under parity it holds $C_{AB}^{\downarrow}(t)=C_{BA}^{\downarrow}(t)$, which combined with
stationarity and the Hermiticity of the matter operators yields $C_{AB}^{\downarrow}(-t)=[C_{AB}^{\downarrow}(t)]^*$, so that the cross kernel \eqref{eq:WAB_finite_time} is real for all $T$. $W_{ab}^{\downarrow}(T;d_{ab})$ samples the correlator over a frequency band of width $\sim1/T$ around $\omega_q$, and in the long-time golden-rule limit it collapses onto the cross spectrum at $\omega_q$. Equation~\eqref{eq:Bell_contrast_second_order} follows directly from the second-order transition probability. The initial Bell state is orthogonal to $|gg\rangle$, so the leading ground-state population comes from the squared first-order transition amplitude. Its interference term is $\Re[e^{-i\phi}W_{BA}^{\downarrow}]$, while the two local probabilities are phase independent and cancel in the half-difference. The secular approximation enters the population resummation below.

\paragraph*{Relaxation signal.}

We initially neglect dephasing and assume that upward transitions
satisfy the finite-time smallness condition
\eqref{eq:Bell_upward_condition} in Appendix~\ref{app:bell}. We retain the probe-frequency-secular relaxation terms, including
resonant cross relaxation between equivalent probes, while keeping
their full finite-time memory integrals. The secular approximation assumes the time-scale separation
specified in Eq.~\eqref{eq:Bell_secular_window}; the additional
condition for neglecting upward transitions is given in
Appendix~\ref{app:bell}. For two equivalent probes, the local relaxation kernels are equal,
\begin{equation}
W_{AA}^{\downarrow}=W_{BB}^{\downarrow}\equiv\overline W.
\label{eq:Wbar}
\end{equation}
The symmetric and antisymmetric Bell states $(|eg\rangle\pm|ge\rangle)/\sqrt2$ then have survival probabilities
$e^{-[\overline W\pm W_{AB}^{\downarrow}]}$
(see Appendix~\ref{app:bell} for a detailed derivation). The preparation phase $\phi$ determines their relative weights\cite{PhysRevA.2.889}, giving the ground-state population
\begin{align}
P_{gg}^{(\phi)}(T,d)={}&1-e^{-\overline W(T)}\Big[\cosh W_{AB}^{\downarrow}(T;d)\nonumber\\
&-
\cos\phi\,\sinh W_{AB}^{\downarrow}(T;d)\Big].
\label{eq:Bell_population_exact}
\end{align}
Comparing opposite Bell phases therefore gives
\begin{equation}
\Delta P_{gg}^{(\phi)}(T,d)
=
e^{-\overline W(T)}
\cos\phi\,
\sinh W_{AB}^{\downarrow}(T;d).
\label{eq:Bell_phase_difference_exact}
\end{equation}
At $\phi=0$, the signal has the same sign as $W_{AB}^{\downarrow}$. It is positive when the symmetric state decays faster and negative when the antisymmetric state decays faster.

At large separation, where $|W_{AB}^{\downarrow}|\ll1$, expanding only the cross contribution gives
\begin{equation}
\Delta P_{gg}^{(\phi)}(T,d)\simeq e^{-\overline W(T)} \cos\phi\, W_{AB}^{\downarrow}(T;d).
\label{eq:Bell_linear_cross}
\end{equation}
When local relaxation is independent of the probe separation, the distance dependence of the signal is set by $W_{AB}^{\downarrow}$, while the factor $e^{-\overline W}$ reduces its amplitude.
\paragraph*{Effect of local dephasing.}
Local dephasing does not directly change the populations, but it weakens the coherence between $|eg\rangle$ and $|ge\rangle$ needed to distinguish opposite Bell phases. Under free precession, this coherence decays with the Ramsey exponent $\chi_R(T)=\chi_{AA}(T)+\chi_{BB}(T)$, obtained from Eq.~\eqref{eq:chi_ab} with $y_z^{(a)}=1$. Keeping the local decay exponents unexpanded and retaining only the leading cross contribution gives
\begin{equation}
\Delta P_{gg}^{(\phi)}(T,d)
\simeq
e^{-\overline W(T)}\cos\phi
\int_0^T dt\,e^{-\chi_R(t)}
\frac{\partial W_{AB}^{\downarrow}(t;d)}{\partial t}.
\label{eq:Bell_linear_dephasing}
\end{equation}
The additional factor describes the loss of phase contrast and tends to unity as $\chi_R\to0$, recovering Eq.~\eqref{eq:Bell_linear_cross} (see Appendix~\ref{app:bell}).

For defect-spin probes with $T_2^*\ll T_1$, dephasing can suppress the signal well before local relaxation becomes appreciable \cite{RevModPhys.92.015004}. Synchronized $\pi$ pulses on both probes can reduce this suppression. With such control, the free-precession kernel \eqref{eq:WAB_finite_time} must be replaced by its control-dressed counterpart constructed from Eq.~\eqref{eq:KAB_def}, and the readout projector $\Pi_{gg}$ by $\widetilde\Pi_{gg}(T)$.

\subsection{Two-qubit dynamical-decoupling covariance}
\label{sec:XY8}

\subsubsection{Protocol and covariance signal}

We specialize the second-order two-qubit dynamics derived above to a covariance protocol following the sensing scheme of Ref.~\cite{Rovny2022}, generalized here to a microscopic quantum environment with arbitrary probe--matter coupling. Dynamical-decoupling pulse sequences act as tunable frequency filters for environmental fluctuations \cite{Green_2013}. We consider two standard implementations: the Carr--Purcell--Meiboom--Gill (CPMG) sequence, which repeats pulses about a single transverse axis, and XY8, which combines pulses about two perpendicular axes to reduce sensitivity to pulse errors. 

Both probes are initially prepared along $+x$,
\begin{equation}
\rho_{AB}(0)=|{+x},{+x}\rangle\langle{+x},{+x}|.
\end{equation}
During the sensing interval $0\leq t\leq T$, each probe undergoes an independently controlled dynamical-decoupling sequence. Probe $a$ receives $N_a$ ideal $\pi$ pulses about the $x$ or $y$ axis, equally spaced by $\tau_a$ and centered on the sensing window, with initial and final free intervals of duration $\tau_a/2$. At the end of the sensing interval, an additional $\pi/2$ readout pulse converts the accumulated phase into a population difference measured in the $z$ basis. Its phase is referenced to the known probe evolution so that, in the interaction picture, the readout rotation is $U_{\mathrm r}^{(a)}=\exp[-i(\pi/2)S_a^x]$. The subsequent $z$-basis measurement is therefore equivalent to measuring
\begin{equation}
\overline M_a = U_{\mathrm r}^{(a)\dagger}S_a^z U_{\mathrm r}^{(a)} = S_a^y =\frac{\sigma_a^y}{2}
\end{equation}
immediately before the readout pulse.

Correlating the two probe outcomes from each experimental repetition
gives the connected covariance. Using Eq.~\eqref{eq:exp_form} with
the block decomposition~\eqref{eq:blocks}, we obtain
\begin{align}
\mathrm{Cov}_{\XY}(T)={}&\Tr\!\big[\overline M_A\overline M_Be^{\Phi_2^{AB}(T)}\rho_{AB}(0)\big]\nonumber\\
&-\Tr\!\big[\overline M_Ae^{\Phi_2^{AB}(T)}\rho_{AB}(0)\big]\nonumber\\
&\quad\times\Tr\!\big[\overline M_Be^{\Phi_2^{AB}(T)}\rho_{AB}(0)\big]+O(V^3).
\label{eq:cov_def}
\end{align}

Ideal $\pi_x$ and $\pi_y$ pulses, with axes defined in the rotating frame, change the signs of the spin components perpendicular to their rotation axes (see Appendix~\ref{app:toggling}). Their pulse-only control matrix is
\begin{equation}
R_{\mathrm{tog}}^{(a)}(t)=\diag\big(y_x^{(a)}(t),y_y^{(a)}(t),y_z^{(a)}(t)\big),
\end{equation}
where $y_\mu^{(a)}(t)=\pm1$ are switching functions. The complete interaction-picture control matrix also includes free precession. For $H_{Q,0}^{(a)}=\omega_q S_a^z$, free precession leaves $S_a^z$ unchanged, so the longitudinal dressed cross-kernel reduces to
\begin{equation}
\mathcal K_{AB}^{zz}(t_1,t_2)=y_z^{(A)}(t_1)y_z^{(B)}(t_2)C_{AB}^{zz}(t_1,t_2).
\end{equation}
The transverse and mixed channels follow from Eq.~\eqref{eq:KAB_def} using the complete control matrices.
 
The XY8 sequence $(\pi_x,\pi_y,\pi_x,\pi_y,\pi_y,\pi_x,\pi_y,\pi_x)$, repeated $M_a$ times ($N_a=8M_a$), modulates all three spin components \cite{Rovny2022}. For CPMG with $\pi_x$ pulses, $y_x^{(a)}=1$ and $y_y^{(a)}=y_z^{(a)}$. The two sequences thus have identical longitudinal filters when their pulse timings and sensing windows coincide, while their transverse filters differ. The longitudinal covariance expressions derived below apply to either sequence when the transverse contributions are negligible. 

The pulse sequence determines which environmental frequencies contribute to the measured covariance. To make this frequency selectivity explicit, we introduce the filter functions
\begin{equation}
F_\mu^{(a)}(\omega)=\int_0^T dt\,y_\mu^{(a)}(t)e^{i\omega t},
\end{equation}
with explicit XY8 expressions given in Appendix~\ref{app:filters}. For otherwise identical sequences, we introduce a relative delay $t_d$ on probe $B$,
\begin{equation}
y_\mu^{(B)}(t;t_d)=y_\mu^{(A)}(t-t_d).
\end{equation}
Neglecting endpoint corrections associated with the shifted measurement window gives
\begin{equation}
F_\mu^{(B)}(\omega;t_d)\simeq e^{i\omega t_d}F_\mu^{(A)}(\omega).
\label{eq:delay_phase}
\end{equation}
Relative filter phases provide sensitivity to the phase of the cross spectrum \cite{PhysRevA.94.012109}. Delayed XY8 sequences have also been used experimentally on independently controlled NV sensors to resolve the temporal structure of two-point field correlations \cite{Rovny2022}. 
\subsubsection{Principal-harmonic approximation}

To obtain a frequency-selective dephasing measurement, we retain the dominant longitudinal response near the principal pulse resonance. For XY8, the transverse filter components vanish at $\omega_\star\simeq\pi/\tau$,
\begin{equation}
f_x^{\XY}(\omega_\star)=f_y^{\XY}(\omega_\star)=0,
\end{equation}
while $f_z^{\XY}(\omega_\star)$ is maximal. We assume that the spectral contributions from higher harmonics and the finite-width transverse filters are negligible. For CPMG, the $y$ and $z$ channels share the same switching function, so this approximation additionally requires negligible matter spectral weight in the corresponding free-precession window.

Under these conditions, the effective interaction is
\begin{equation}
\widetilde V_{\XY}(t)
\simeq
\sum_{a=A,B}
y_z^{(a)}(t)S_a^z
\otimes\delta\widetilde{\mathcal O}_a^z(t).
\label{eq:Veff_XY8}
\end{equation}
This interaction preserves the populations in the $S^z$ basis and changes only the coherences between its states. In the product basis of $S^z_A$ and $S^z_B$ the second-order generator then acts diagonally, each coherence $|m\rangle\langle n|$ decaying with its own real exponent. Since the readout operators are $S^y_A$ and $S^y_B$, the joint signal involves only the two coherences between $|00\rangle$ and $|11\rangle$ and between $|01\rangle$ and $|10\rangle$, for which the two basis states have opposite $S^z$ eigenvalues at each probe. The response part of the generator does not contribute to these two coherences, and their exponents are built from the symmetrized correlation functions $N^{zz}_{ab}$ alone, through the local and cross dephasing exponents
\begin{equation}
\chi_{ab}(T,t_d) =\half \int_0^T dt_1 \int_0^T dt_2 y_z^{(a)}(t_1) y_z^{(b)}(t_2;t_d) N_{ab}^{zz}(t_1,t_2).
\label{eq:chi_ab}
\end{equation}
The local exponents describe the dephasing of each probe, while $\chi_{AB}$ describes the contribution of the correlations between the sensing regions \cite{PhysRevA.94.012109,Paz-Silva_2016}. For the initial state $|{+x},{+x}\rangle$ both coherences start at $1/4$, and for the symmetric preparation and pulse protocol considered here the single-probe expectation values $\langle S^y_A\rangle$ and $\langle S^y_B\rangle$ vanish, so that the joint signal equals the connected covariance. Collecting these elements we find (Appendix~\ref{app:cov}),
\begin{equation}
\mathrm{Cov}_{\XY}(T,t_d)=\frac{1}{4}e^{-[\chi_{AA}(T)+\chi_{BB}(T)]}\sinh\big[2\chi_{AB}(T,t_d)\big].
\label{eq:CovarianceXY8}
\end{equation}

Equation~\eqref{eq:CovarianceXY8} generalizes the covariance-magnetometry result derived for classical correlated fields in Ref.~\cite{Rovny2022} to an arbitrary quantum many-body environment. Here $\chi_{AB}$ is fixed by the nonlocal symmetrized correlation function of the matter operators sampled by the two probes, dressed by the two control filters. In the Ramsey limit of free precession, $y_z^{(a)}=1$, the exponents reduce to unfiltered double integrals of the correlators, and we recover the correlated-Ramsey result of Ref.~\cite{Hosseinabadi2026} for two qubits prepared along $x$. In that case the filter $|F_z(\omega)|^2=4\sin^2(\omega T/2)/\omega^2$ is centered at zero frequency, and the measurement is sensitive to the slowest fluctuations of the environment, independent of the sensing time. With the $\pi$-pulse sequence used here, the filter is instead centered at $\omega_\star=\pi/\tau$, and the measurement becomes sensitive to the fluctuations at one chosen frequency, which is moved across the spectrum by varying the pulse spacing $\tau$. This is what makes the covariance a frequency-resolved probe, explored in detail in Sec.~\ref{sec:XY8_LL}.

Further simplifications arise from the stationarity of the environment, Eq.~\eqref{eq:init}, under which the correlator depends only on the time difference, $N^{zz}_{AB}(t_1,t_2)=N^{zz}_{AB}(t_1-t_2;d)$, and the double time integral in Eq.~\eqref{eq:chi_ab} factorizes into the two filter functions,
\begin{equation}
\chi_{AB}(T,t_d) =\half\int_{-\infty}^{\infty}\frac{d\omega}{2\pi} e^{-i\omega t_d}
|F_z(\omega)|^2 \overline S_{AB}^{zz}(\omega;d),
\label{eq:chiAB_freq}
\end{equation}
with $\overline S_{AB}^{zz}(\omega;d)=\int_{-\infty}^{\infty}d\tau e^{-i\omega\tau} N_{AB}^{zz}(\tau;d)$. 

The delay dependence is further constrained by symmetry. For equivalent probes in a reflection-symmetric environment, the longitudinal correlator is invariant under exchange of the probe sites. For identical pulse trains, this exchange reverses $t_d$. With the shifted-window convention of Eq.~\eqref{eq:delay_phase}, the longitudinal covariance in Eq.~\eqref{eq:CovarianceXY8} therefore obeys
\begin{equation}
\mathrm{Cov}_{\XY}(T,t_d) =\mathrm{Cov}_{\XY}(T,-t_d).
\label{eq:cov_even}
\end{equation}
Extending this symmetry argument to the full tensorial covariance requires the exchange operation to preserve the preparation, controls, and readout. At the level of the kernels the same symmetry gives $N_{AB}^{zz}(\tau;d)=N_{BA}^{zz}(\tau;d)=N_{AB}^{zz}(-\tau;d)$, so that $\overline S_{AB}^{zz}(\omega;d)$ is real and even in $\omega$, and Eq.~\eqref{eq:chiAB_freq} reduces to
\begin{equation}
\chi_{AB}(T,t_d) = \half\int \frac{d\omega}{2\pi}\,\cos(\omega t_d)\,|F_z(\omega)|^2\overline S_{AB}^{zz}(\omega;d).
\label{eq:chiAB_even}
\end{equation}
The cross exponent is thus real and even in the delay, and the propagation of correlations between the probes cannot appear as a phase. Finally, when many pulses fit in the sensing window, $N_a\gg1$, the filter is concentrated at $\pm\omega_\star$ within a width of order $1/T$ \cite{PhysRevLett.107.230501}, and for $t_d\ll T$ Eq.~\eqref{eq:chiAB_even} factorizes,
\begin{equation}
\chi_{AB}(T,t_d) \simeq \mathcal W_{\XY}(T)\, \cos(\omega_\star t_d)\, \overline S_{AB}^{zz}(\omega_\star;d)\,,
\label{eq:chiAB_factorized}
\end{equation}
where $\mathcal W_{\XY}(T)$ is the integrated weight of the two principal filter peaks defined in Eq.~\eqref{eq:app_XY8_weight}. After the factorization, the propagation of correlations is encoded in the dependence of $\overline S_{AB}^{zz}(\omega_\star;d)$ on the separation at fixed filter frequency, or on the filter frequency at fixed separation.

\section{Application to an interacting quantum spin chain}
\label{sec:spinchain}

We now specialize the general framework to a microscopic interacting spin system. The quantum matter is a spin-$1/2$ XXZ chain, one of the most fundamental strongly correlated quantum systems and a famous example of a gapless algebraic quantum spin liquid, hosting complex quantum many-body physics \cite{Breunigeaao3773,PhysRevLett.111.187202,Psaroudaki_2016}. The model hosts rich interaction-dependent noncommutative algebraic structures \cite{KITANINE1999647,Caux_2005}, which provide a natural setting for testing two-probe sensing of quantum many-body dynamics. The two probes couple to the quantum chain through a coupling of dipolar form with spatial range $h$. The dipolar form reflects the stray-field coupling of a magnetic sensor, and any coupling of range $h$ acts in the same way, as a momentum filter of width $1/h$. We also consider a single-site exchange coupling of an impurity qubit in a quantum simulator, introduced as a separate coupling model. This makes the relation between the generic cross-kernels introduced above and the intrinsic spatiotemporal spin correlations of the many-body system explicit.

\subsection{XXZ model and dipolar coupling}
\label{sec:XXZ_model}

We consider the many-body Hamiltonian of the anisotropic $S=1/2$ Heisenberg AFM chain
\begin{equation}
H_M=J\sum_{n=1}^{N-1}\left(s_n^x s_{n+1}^x+s_n^y s_{n+1}^y+\Delta s_n^z s_{n+1}^z\right)-h_z\sum_{n=1}^{N}s_n^z,
\label{eq:HB}
\end{equation}

where $s_n^{a}=\hbar \sigma_n^a /2$, with $\sigma_n^a$ the Pauli matrices, $N$ the number of sites, $J>0$ is the AFM interaction strength, and $\Delta$ the anisotropy parameter. At the isotropic point $\Delta=1$, the model is realized in several one-dimensional AFM materials including Sr$_2$CuO$_3$, SrCuO$_2$, and CaCu$_2$O$_3$ crystals \cite{PhysRevLett.76.3212,HESS20191}. $\Delta=1/2$ also serves as a low-energy effective model for the magnetic-field induced gapless phase of $S=1$ NiCl$_2$ -SC(NH$_2$ )$_2$ \cite{PhysRevB.89.224418}, while a spin chain with arbitrary $J$ and $\Delta$ can be realized in an array of laser-trapped circular Rydberg atoms with dynamically tunable model parameters \cite{PhysRevX.8.011032}. 

\begin{figure*}[t]
\centering
\includegraphics[width=0.9\linewidth]{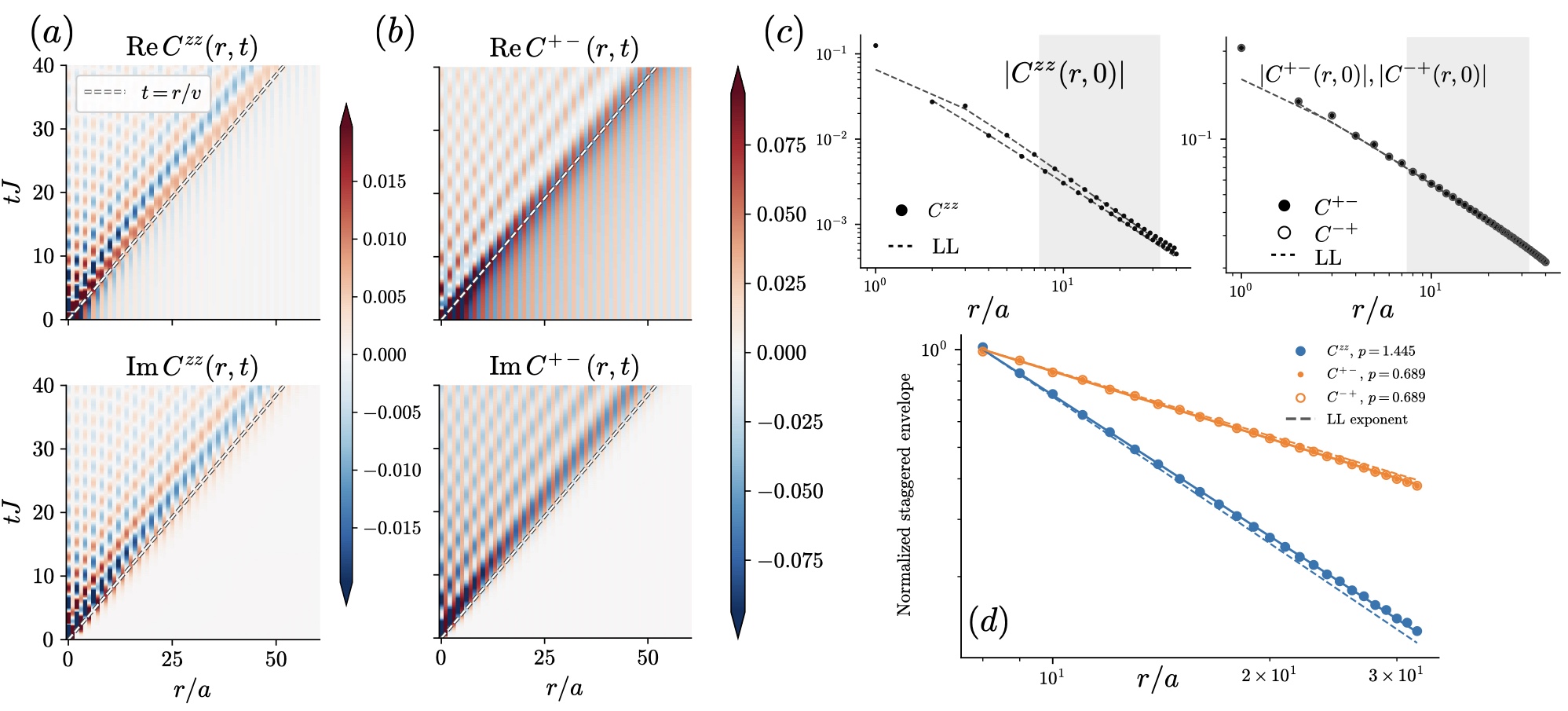}
\caption{\textbf{Spin correlations of the XXZ chain.} (a,b) Space--time maps of the longitudinal correlator $C^{zz}(r,t)$ and the transverse correlator $C^{+-}(r,t)$, respectively, for $h_z=0$ and $\Delta=1/2$. The upper and lower rows show the real and imaginary parts. The dashed line marks $t=r/v$, with the Bethe-ansatz spin velocity $v/(Ja)=1.299$. The real transverse correlator retains the staggered ground-state correlations ahead of the front, while the imaginary part highlights the propagating response. (c) Equal-time magnitudes. The dashed curves are Luttinger-liquid predictions at $K=3/4$, with amplitudes fitted over the shaded interval $8\leq r/a\leq32$. The two transverse channels coincide by spin-flip symmetry. (d) Staggered envelopes over the same interval, with the smooth $r^{-2}$ term removed from $C^{zz}$ and each envelope normalized to its fitted value at $r=8a$. The $r^{-p}$ fits (solid) give $p=1.445$ longitudinally and $0.689$ transversely, compared with the LL exponents $2K=3/2$ and $1/(2K)=2/3$ (dashed).}
\label{fig:Spin_Corr}
\end{figure*}

The generic operators $\mathcal O_a^\alpha$ of Eq.~\eqref{eq:Va_def} become the magnetic-field operators $B_a^\alpha=\sum_{j,\beta}D_{a,j}^{\alpha\beta}s_j^\beta$
and $\delta B_a^\alpha=B_a^\alpha-\langle B_a^\alpha\rangle_B$. The probe--chain interaction is therefore $V=V_A+V_B$ with 
\begin{equation}
V_a=\sum_{\alpha=x,y,z}S_a^\alpha\otimes\delta B_a^\alpha.
\label{eq:Va_dipolar}
\end{equation}
The dipolar tensor is
\begin{equation}
D_{a,j}^{\alpha\beta} =\frac{\mu_0\gamma_Q\gamma_s}{4\pi|\mathbf r_{a,j}|^3}
\left(\delta_{\alpha\beta}-3\hat r_{a,j\alpha}\hat r_{a,j\beta}\right),
\label{eq:dipolar_tensor}
\end{equation}
with $\mathbf r_{a,j}=\mathbf R_a-\mathbf R_j$. The mean-field part is absorbed into the renormalized qubit Hamiltonian, and only the $\delta s_j^\beta$ fluctuations appear. We choose to work in the coplanar geometry
\begin{equation}
\mathbf R_j=(ja,0,0),\quad\mathbf R_A=(0,0,h),\quad\mathbf R_B=(d,0,h),
\label{eq:geometry}
\end{equation}
so that $h$ is the probe--sample distance and $d=|\mathbf R_A-\mathbf R_B|$ is the probe separation along the chain. The separation enters the microscopic coupling only through the relative shift of the two dipolar tensors $D_{A,j}^{\alpha\beta}$ and $D_{B,j}^{\alpha\beta}$.

The generic two-probe correlator Eq.~\eqref{eq:Cab_def} now becomes
\begin{align}
C_{ab}^{\alpha\beta}(t_1,t_2)=\sum_{j,j'}\sum_{\gamma,\delta}D_{a,j}^{\alpha\gamma}D_{b,j'}^{\beta\delta}\langle\delta\widetilde s_j^\gamma(t_1)\delta\widetilde s_{j'}^\delta(t_2)\rangle_B. 
\label{eq:Cab_micro}
\end{align}
The local kernels $C_{AA}$ and $C_{BB}$ involve autocorrelations of the chain weighted by each probe's form factor, whereas the cross-kernels $C_{AB}$ and $C_{BA}$ involve \emph{nonlocal} spatiotemporal spin correlations between the neighborhoods sampled by the two probes. For a stationary, translationally invariant chain state, it also holds $C_{ab}^{\alpha\beta}(t_1,t_2)=C_{ab}^{\alpha\beta}(t_1-t_2;d_{ab})$, with $d_{AA}=d_{BB}=0$ and $d_{AB}=d_{BA}=d$. Such off-diagonal bath correlators are the microscopic origin of spatially correlated relaxation and dephasing in multi-qubit open-system descriptions \cite{Gulacsi2025,Zou2024}.

For arbitrary $\Delta$ the bath Hamiltonian in a longitudinal field conserves
total magnetization, $S_{\rm tot}^z =\sum_n s_n^z$ \cite{PhysRevLett.95.077201}. If the bath state also satisfies $[\rho_B,S_{\rm tot}^z]=0$, only $U(1)$ two-point functions survive,
\begin{equation}
\langle s_j^\mu(t_1)s_{j'}^\nu(t_2)\rangle
=
\delta_{q_\mu+q_\nu,0}\,
\langle s_j^\mu(t_1)s_{j'}^\nu(t_2)\rangle,
\label{eq:neutral}
\end{equation}
where $\mu,\nu\in\{z,+,-\}$, with $q_z=0$ and $q_\pm=\pm1$.
The two-qubit kernels
\eqref{eq:KAB_def}
therefore reduce to longitudinal and transverse sectors only, and the cross-kernel $C_{AB}$ is built from exactly the same dynamical spin correlators as the single-probe problem, sampled at finite spatial separation $d$. Only at zero field and for an $SU(2)$-invariant state does the further simplification $C_{jj'}^{\alpha\beta}=\delta_{\alpha\beta}C_{jj'}$ hold.

In the stationary, translation-invariant limit, it is natural to introduce the
momentum--frequency representation
\begin{equation}
C^{\beta\beta'}(r,\tau)=\int_{-\pi/a}^{\pi/a}
\frac{dq}{2\pi}\int_{-\infty}^{\infty}\frac{d\omega}{2\pi}\,e^{iqr-i\omega\tau}
\overline S^{\beta\beta'}(q,\omega),
\label{eq:Sqomega}
\end{equation}
where
$\overline S^{\beta\beta'}(q,\omega)$ is the dynamical spin structure factor. In this representation, the spatially extended dipolar coupling enters through a momentum-space form factor and the control through $R^{(a)}(t)$.

For the geometry \eqref{eq:geometry}, the relevant components are
\begin{equation}
\bigl(D^{zz},D^{zx}\bigr)=\frac{g_{\rm dip}\bigl(x^2-2h^2,-3xh\bigr)}
{(x^2+h^2)^{5/2}}
\label{eq:Dzz_Dzx}
\end{equation}
and $D^{zy}=0$, and $D^{z\pm}=\tfrac12D^{zx}$. Their Fourier transforms decay exponentially at large momentum,
\begin{equation}
|D^{zz}(q;h)|^2,\;|D^{zx}(q;h)|^2\propto e^{-2|q|h},
\label{eq:formfactor}
\end{equation}
so that the probe height acts as a momentum-space filter suppressing fluctuations with $|q|h\gg1$. This wave-vector selectivity of quantum-impurity relaxometry is well established \cite{PhysRevLett.121.187204,PhysRevB.98.195433}, enters two-qubit dephasing spectroscopy of a magnetic film in the same way \cite{Hosseinabadi2026}, and is particularly consequential for lattice-scale correlations. When $h\gg a$, staggered fluctuations are exponentially suppressed and the probe predominantly samples smooth long-wavelength sectors \cite{PhysRevLett.121.187204,PhysRevLett.131.070801}. As we show below, this geometric filter, combined with the intrinsic scaling dimensions of the quantum chain, determines which Luttinger-liquid sector dominates the two-probe signal.

In the gapless phase, the spin-$1/2$ XXZ chain is described at low energies by a Tomonaga--Luttinger liquid (LL), a Gaussian theory of collective bosonic modes \cite{PhysRevLett.45.1358}. Its universal behavior is governed by the excitation velocity $v$ and the dimensionless Luttinger parameter $K$. Both depend on the exchange anisotropy and magnetization and can be obtained from the Bethe ansatz. At zero temperature, the bulk equal-time longitudinal correlator contains a smooth $r^{-2}$ term and an oscillating contribution proportional to $\cos(2k_F r)\,r^{-2K}$. Here $k_F$ is the Fermi wave vector of the Jordan--Wigner fermions, with $2k_F=\pi(1-2m)/a$, where $m=\langle S_{\mathrm{tot}}^z\rangle/L$ is the magnetization per site and $a$ is the lattice spacing. At zero magnetization, $2k_F=\pi/a$ and the oscillating term is staggered. The leading transverse correlations are staggered and decay as $r^{-1/(2K)}$ \cite{LUKYANOV2003323,PhysRevB.69.064427}.

We compute the correlators numerically using tensor-network methods implemented in TeNPy~\cite{tenpy2024}. For open chains with $\Delta=1/2$, we obtain the ground state by DMRG, apply a spin operator at the central site, and evolve the resulting matrix product state up to $tJ=40$ using fourth-order time-evolving block decimation (TEBD). We use $L=140$ at zero field and $L=141$ at $m\simeq0.202$. See Appendix~\ref{app:numerics} for details. The zero-field data in Fig.~\ref{fig:Spin_Corr} (a,b) show a ballistic front near $t=r/v$. The equal-time data and fits in Fig.~\ref{fig:Spin_Corr}(c,d) show the slower decay of the staggered transverse correlations, in agreement with the LL predictions.

These power laws reflect quasi-long-range order throughout an extended quantum-critical phase. Conventional short-range classical spin chains instead have ordered ground states and a finite correlation length at nonzero temperature \cite{10.1119/1.1970340}. Algebraic decay also occurs in the XX limit ($\Delta=0$), which maps to noninteracting fermions and has $K=1$, giving longitudinal and transverse exponents of $2$ and $1/2$, respectively. Interactions renormalize $K$ continuously, changing the two exponents in complementary ways \cite{PhysRevLett.45.1358,PhysRevB.69.064427}. Their joint measurement therefore probes how interactions shape the spatial correlations sampled by the two-probe protocols.

\begin{figure*}[t]
\centering
\includegraphics[width=1\linewidth]{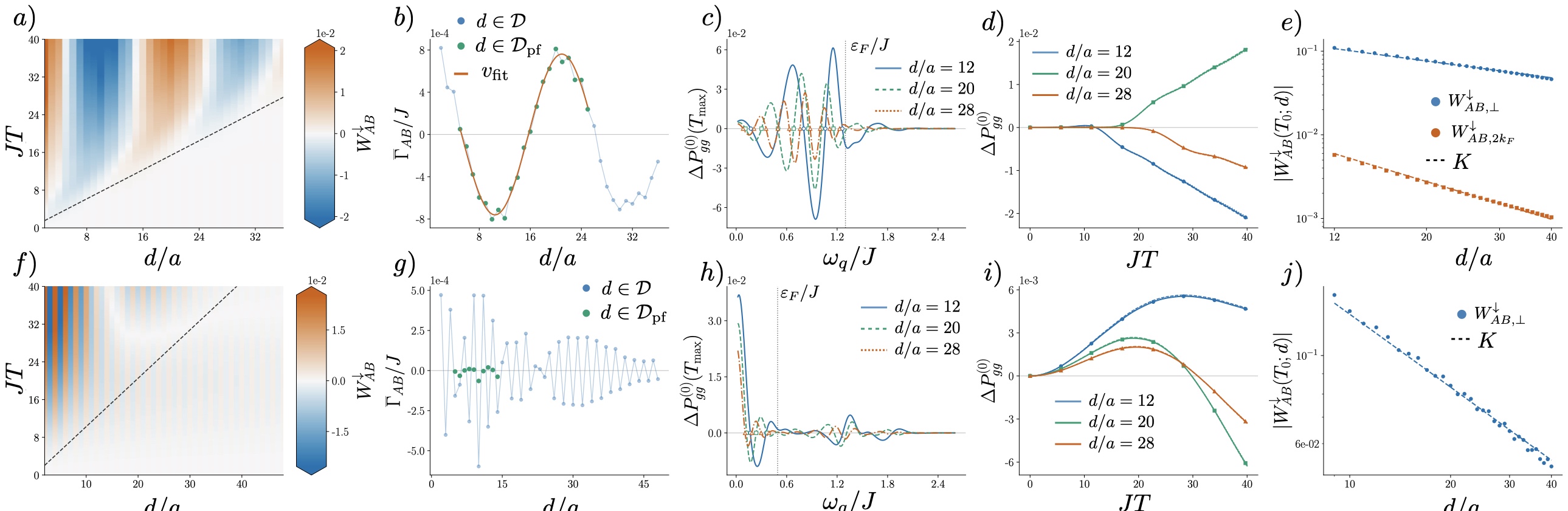}
\caption{\textbf{Bell-state relaxometry}. Bell-state relaxometry results for an XXZ chain at $\Delta=1/2$. The top row uses $L=140$ at zero field. The bottom row uses $L=141$ in the sector $m=28.5/141\simeq0.202$, with the transverse correlators rephased from $h_z/J=0.96$ to $h_z/J\simeq0.970$. For the dipolar panels, the probe heights are $h/a=2$ and $1$, respectively. (a,f) Downward cross-relaxation kernel $W_{AB}^{\downarrow}(T;d)$. The dashed line marks the ballistic front $d=v_{\mathrm{BA}}T$, with $v_{\mathrm{BA}}$ the Bethe--ansatz velocity.
(b,g) Cross-relaxation rate $\Gamma_{AB}=\partial_T W_{AB}^{\downarrow}$ averaged over the final $4/J$ of the simulation. Pale markers show all separations $d\in\mathcal D$, and green markers select the post-front window $\mathcal D_{\mathrm{pf}}$, defined by $d\geq 5a$ and $T_{\max}-4/J\geq d/v_{\mathrm{BA}}+2\pi/\omega_q$. The orange curve fit gives $v_{\mathrm{fit}}/(Ja)\simeq 1.30$. The green points in (g) use a three-site average to reduce the even--odd modulation. (c,h) Signed Bell contrast $\Delta P_{gg}^{(0)}(T_{\max})$ versus qubit frequency $\omega_q/J$, computed in the downward-only approximation at $JT_{\max}=40$. The vertical dotted lines mark $\varepsilon_F/J$, and open symbols indicate resolved zero crossings below $\varepsilon_F$. (d,i) Signed Bell contrast versus sensing time at $d/a=12,20,28$. Solid curves with markers show the full reduced-map calculation. The downward-only and local-dephasing approximations are shown by dashed and dotted curves, respectively, and nearly overlap the full result. The fixed qubit frequencies are $\omega_q/J=0.39$ in (a,b,d), $0.15$ in (f,i), and $0.30$ in (g). (e,j) Magnitudes of the single-site kernels at $JT_0=3$ and $\omega_q/J=0.25$. Blue circles show the staggered transverse contribution $W_{AB,\perp}^{\downarrow}$, and orange squares in (e) show the longitudinal $2k_F$ contribution $W_{AB,2k_F}^{\downarrow}$. The dashed lines have the Bethe--ansatz power laws $d^{-1/(2K)}$ and $d^{-2K}$, respectively, with only their amplitudes fitted to the numerical points.}
\label{fig:Bell_relaxometry}
\end{figure*}
\subsection{Bell-state relaxometry in the XXZ chain}
\label{sec:Bell_XXZ}

For the full dipolar coupling, the transverse qubit bath operator appearing in the generic relaxation kernel \eqref{eq:WAB_finite_time} contains both longitudinal and transverse chain-spin sectors. Using $\delta B_a^+=\delta B_a^x+i\delta B_a^y$ gives
\begin{equation}
\delta B_a^+=\sum_j\left(D_{a,j}^{+z}\delta s_j^z+D_{a,j}^{++}\delta s_j^+
+D_{a,j}^{+-}\delta s_j^-\right),
\label{eq:Bplus_spin_decomp}
\end{equation}
where $D^{+z}=D^{xz}+iD^{yz}$ and $D^{+\pm}=1/2\left[D^{xx}\pm D^{yy} +i(D^{yx}\mp D^{xy})\right]$. 
We denote by $\mathcal D_{zz}(q;h)$ and $\mathcal D_\perp(q;h)$ the Fourier-space dipolar coefficients multiplying the longitudinal and dominant transverse chain sectors in $B^+$, respectively. By the $U(1)$ selection rules \eqref{eq:neutral}, the Bell signal therefore contains the full dipolar-weighted nonlocal $zz$, $+-$, and $-+$ correlations of the chain.

\subsubsection{Luttinger-liquid predictions for Bell relaxometry}
\label{sec:Bell_LL}

At zero temperature, the sensing time determines which part of the gapless XXZ chain's correlation dynamics enters the Bell cross kernel. For probes
separated by $d$, the characteristic time is $d/v$, where $v$ is the Luttinger-liquid velocity. At times $T\ll d/v$, the kernel mainly samples the
spatial correlations already present in the ground state. When $T$ reaches $d/v$, the sensing window includes the ballistic correlation front. At longer times, the kernel resolves the bath spectrum near the qubit frequency $\omega_q$. We first describe this crossover for the two oscillating sectors, then discuss the smooth longitudinal contribution and the role of probe height.

\paragraph*{Oscillating sectors.}
The longitudinal $2k_F$ and staggered transverse sectors have wave numbers $Q_x=2k_F$ and $\pi/a$ and scaling dimensions $x=K$ and $1/(4K)$, respectively [Eq.~\eqref{eq:LL_sectors}]. At $d\gg a$, their zero-temperature Wightman correlators take the common form
\begin{align}
C_x^{\downarrow}(d,t)
\simeq{}& A_x\cos(Q_xd+\varphi_x)\\
&\times\left[
\frac{a^2}{(d-vt-ia)(d+vt+ia)}
\right]^x ,
\label{eq:LL_Wightman_zeroT}
\end{align}
where $A_x$ is nonuniversal and the lattice spacing $a$ sets the short-distance cutoff. The two factors become large near $t=\pm d/v$. The pole prescription corresponds to the operator ordering in Eq.~\eqref{eq:Cdown_generic}, with the time-dependent operator on the right.
It selects positive frequencies under the transform $e^{-i\omega_q t}$ used for downward probe transitions.

If the dipolar form factor is approximately constant across the momenta sampled near $Q_x$ (a width of order $\omega_q/v$ in the frequency-resolved regime), inserting Eq.~\eqref{eq:LL_Wightman_zeroT} into the finite-time kernel [Eq.~\eqref{eq:WAB_finite_time}] gives
\begin{align}
W_{AB,x}^{\downarrow}(T;d)
\simeq{}&
\frac{A_x}{4}|\mathcal D_x(Q_x;h)|^2
\left(\frac{a}{d}\right)^{2x}
\left(\frac{d}{v}\right)^2
\nonumber\\
&\times \cos(Q_xd+\varphi_x)\,
\mathcal F_x\!\left(\frac{vT}{d},\frac{\omega_qd}{v}\right).
\label{eq:WAB_scaling_form}
\end{align}
Here $h$ is the probe height and $\mathcal F_x$ is given in Eq.~\eqref{eq:Bell_scaling_function} of Appendix~\ref{app:LL}. The ratio $vT/d$ determines whether the sensing window reaches the front.

\paragraph*{Before the front ($vT\ll d$).}
Over the sensing window, the correlator stays close to its equal-time value. The time integration then contributes only the finite-time filter factor,
\begin{align}
W_{AB,x}^{\downarrow}(T;d)
\simeq{}&
A_x|\mathcal D_x(Q_x;h)|^2
\left(\frac{a}{d}\right)^{2x}
\nonumber\\
&\times\cos(Q_xd+\varphi_x)\,
\frac{\sin^2(\omega_qT/2)}{\omega_q^2}.
\label{eq:Bell_pre_lightcone}
\end{align}
At fixed $T$ and $\omega_q$, the spatial envelopes therefore inherit the equal-time exponents,
\begin{equation}
W_{AB,2k_F}^{\downarrow}\propto d^{-2K},
\qquad
W_{AB,\perp}^{\downarrow}\propto d^{-1/(2K)} .
\label{eq:Bell_precone_exponents}
\end{equation}
The Bell signal can already be finite in this regime because the ground state has spatial correlations. The light cone marks the arrival of a dynamical enhancement in the Wightman correlator. A retarded response probes a different quantity, the commutator, which vanishes outside the light cone in the continuum Luttinger liquid.

\paragraph*{At the front ($vT\simeq d$).}
The triangular time window makes the front especially clear in the curvature of the kernel. Differentiating Eq.~\eqref{eq:WAB_finite_time} gives
\begin{align}
\frac{dW_{AB}^{\downarrow}}{dT}
&=\frac14\int_{-T}^{T}dt\,e^{-i\omega_qt}
C_{AB}^{\downarrow}(d,t),
\label{eq:first_derivative_WAB}\\
\frac{d^2W_{AB}^{\downarrow}}{dT^2}
&=\frac14\bigl[
e^{-i\omega_qT}C_{AB}^{\downarrow}(d,T)
\nonumber\\[-0.2ex]
&\hphantom{=\frac14\bigl[}\,
+e^{i\omega_qT}C_{AB}^{\downarrow}(d,-T)
\bigr].
\label{eq:second_derivative_WAB}
\end{align}
The curvature samples the correlator at the two edges of the sensing window. Tracking the feature near $T=d/v$ across separations gives $v$. In practice, the short-distance cutoff and the time resolution round the feature. A probe at height $h$ also samples a region of order $h$, which gives a geometric broadening $\delta T_{\rm geom}\sim h/v$.

\paragraph*{After the front ($T\gg d/v$, $\omega_qT\gg1$).}
The frequency filter is now narrow enough to resolve the Wightman spectrum near $\omega_q$, so
$W_{AB,x}^{\downarrow}(T;d)\simeq (T/4)S_{AB,x}^{\downarrow}(d,\omega_q)$. The spatial transform of the low-energy structure factor is a Bessel function [Eq.~\eqref{eq:Bell_Bessel}]. For $\omega_qd/v\gg1$ and $\omega_q$ within the Luttinger-liquid regime, its large-distance form is
\begin{align}
W_{AB,x}^{\downarrow}(T;d)
\simeq{}&
\mathcal B_x(T,\omega_q)|\mathcal D_x(Q_x;h)|^2
\left(\frac{a}{d}\right)^x
\left(\frac{\omega_qa}{v}\right)^{x-1}
\nonumber\\
&\times\cos(Q_xd+\varphi_x)
\cos\!\left(\frac{\omega_qd}{v}-\frac{\pi x}{2}\right),
\label{eq:Bell_postcone_asymptotic}
\end{align}
where $\mathcal B_x$ is real and proportional to $T$. For mirror-equivalent probes, the two fronts contribute complex-conjugate terms, giving the real
cosine in Eq.~\eqref{eq:Bell_postcone_asymptotic}, consistent with Eq.~\eqref{eq:parity_Bell}. The fixed-frequency spatial envelopes are
\begin{align}
W_{AB,2k_F}^{\downarrow}
&\propto |\mathcal D_{zz}(2k_F;h)|^2 (a/d)^K,
\nonumber\\
W_{AB,\perp}^{\downarrow}
&\propto |\mathcal D_\perp(\pi/a;h)|^2 (a/d)^{1/(4K)} .
\label{eq:Bell_postcone_envelopes}
\end{align}
The same Bell kernel thus crosses over from the equal-time decay $d^{-2x}$ at short sensing times to the frequency-resolved decay $d^{-x}$ at long
sensing times. At the zero-field isotropic Heisenberg point, $K=1/2$ and both long-time envelopes decay as $d^{-1/2}$, up to multiplicative logarithms.

The last cosine in Eq.~\eqref{eq:Bell_postcone_asymptotic} gives a second way to measure $v$. When a single sector dominates and the bath parameters
are effectively fixed during the sweep, consecutive sign changes of $\Delta P_{gg}^{(0)}$ occur at qubit frequencies separated by
\begin{equation}
\Delta\omega_q \simeq \frac{\pi v}{d}.
\label{eq:Bell_group_delay}
\end{equation}
This spacing measures the same propagation time $d/v$ as the time-domain feature. If the applied field tunes both $\omega_q$ and the chain magnetization, the sweep must account for the field dependence of $v$ and $K$ (Sec.~\ref{sec:exp}).

\paragraph*{Smooth longitudinal sector and probe height.}
The smooth longitudinal spin density is proportional to $-\partial_x\phi$. It is a sum of left- and right-moving currents, with downward correlator
\begin{equation}
C_0^{zz,\downarrow}(d,t)
\simeq A_0\left[
\frac{a^2}{(d-vt-ia)^2}
+\frac{a^2}{(d+vt+ia)^2}
\right].
\label{eq:smooth_chiral_correlator}
\end{equation}
Before the front, this gives a $d^{-2}$ envelope if the projected dipolar weight is nonzero at $q=0$. In the frequency-resolved regime, the ideal linear spectrum lies on $\omega=v|q|$. At fixed $\omega_q$, the two branches therefore give
\begin{equation}
W_{AB,0}^{\downarrow}(T;d)
\simeq \mathcal B_0(T,\omega_q)
|\mathcal D_0(q_q;h)|^2\cos(q_qd),
\label{eq:smooth_ballistic_signal}
\end{equation}
where $q_q=\omega_q/v$. The envelope of the ideal smooth contribution is independent of $d$ in this regime. Spectral broadening beyond the linear theory can attenuate this contribution. When the smooth sector dominates, the small cross contribution assumed in Eq.~\eqref{eq:duhamel} must be checked directly.

As in the XY8 protocol, increasing the height filters out large wave numbers. The leading height dependence of the squared form factor is $e^{-2|Q|h}$, where $Q$ is the sector wave number folded into the first Brillouin zone. Height therefore favors the longitudinal $2k_F$ sector over the staggered transverse sector when its wave number has magnitude below $\pi/a$. When both wave numbers coincide, the exponential height dependence is the same. At larger heights the smooth sector can dominate if the dipolar coupling has appreciable weight at $q_q$.
\subsubsection{Ballistic propagation and spatial scaling in Bell relaxometry}
\label{sec:num_bell}

Figure~\ref{fig:Bell_relaxometry} compares the Bell response at zero magnetization and at $m\simeq0.202$. For the dipolar coupling, the probe heights are $h=2a$ and $h=a$, respectively. Panels (a) and (f) show the finite-time cross-relaxation kernel $W_{AB}^{\downarrow}(T;d)$ of Eq.~\eqref{eq:WAB_finite_time}. The dashed line $d=v_{\mathrm{BA}}T$ marks when the sensing window reaches the ballistic correlation front at the Bethe--ansatz velocity. The pronounced even--odd modulation in panel (f) reflects a staggered contribution to the dipolar kernel.

To examine propagation after the front, we average the rate $\Gamma_{AB}=\partial_T W_{AB}^{\downarrow}$ over the final $4/J$ of the simulation [panels (b) and (g)]. The highlighted separations satisfy $d\geq5a$ and $T_{\max}-4/J\geq d/v_{\mathrm{BA}}+2\pi/\omega_q$, so the averaging window starts at least one qubit period after the front arrives. At $m=0$, fitting $\overline{\Gamma}_{AB}(d)=A\cos(\omega_qd/v_{\mathrm{fit}})$ gives $v_{\mathrm{fit}}/(Ja)\simeq1.30$, close to $v_{\mathrm{BA}}/(Ja)=1.299$. The spatial oscillation therefore provides a velocity estimate through Eq.~\eqref{eq:smooth_ballistic_signal}. At finite magnetization, a three-site average reduces the even--odd modulation. The available post-front separations span too small a range for a comparable fit.
\begin{figure*}[t]
\centering
\includegraphics[width=1\linewidth]{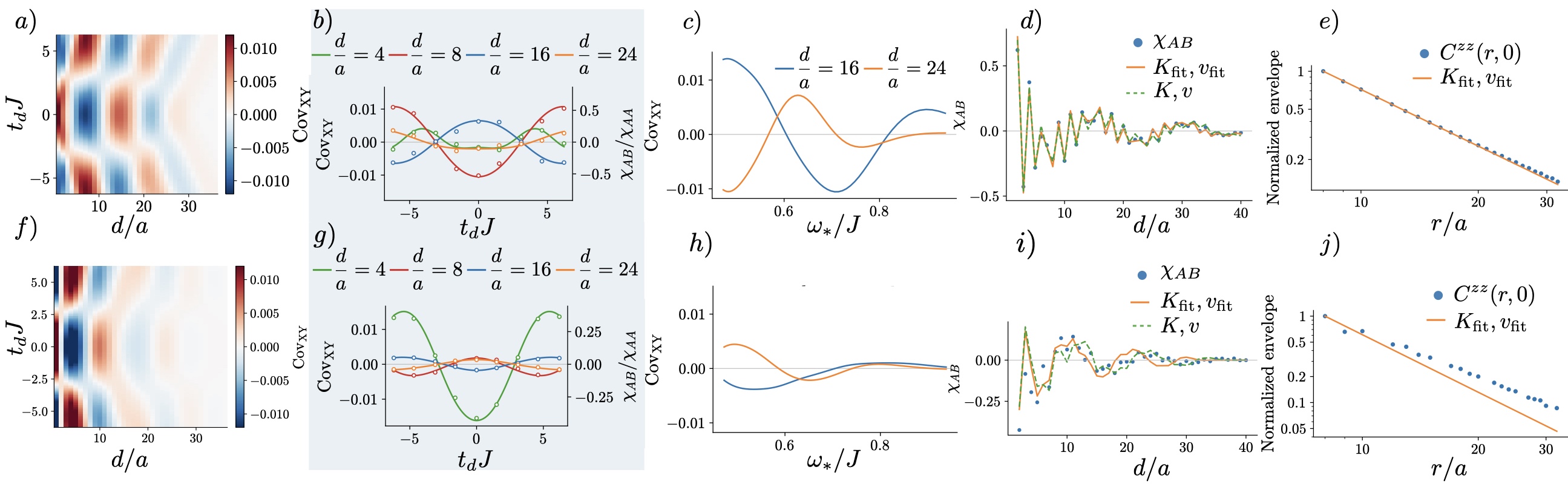}
\caption{\textbf{Two-probe covariance}. Covariance results for an XXZ chain at $\Delta=1/2$. The top row uses $L=140$ at $h_z/J=0$, and the bottom row uses $L=141$ at $h_z/J=0.96$. (a,f) Dipolar covariance $\mathrm{Cov}_{\mathrm{XY}}$ versus probe separation $d/a$ and relative pulse-train delay $t_dJ$, computed with the longitudinal filter for four pulses at $\tau J=6.2$ and probe height $h/a=2$. Both maps use the same color scale. (b,g) Delay cuts through the maps at the indicated separations. Solid curves show the longitudinal-filter result and open circles show the full second-order time-convolutionless (TCL2) calculation. The right axis gives $\chi_{AB}/\chi_{AA}$ for the solid curves. (c,h) Dipolar covariance at zero delay for six pulses, plotted versus the filter frequency $\omega_\star=\pi/\tau$ at the indicated probe separations. The filter frequency is varied by changing the interpulse spacing $\tau$. (d,i) Cross-dephasing $\chi_{AB}$ for probes coupled to single sites, shown at $TJ=37.2$ and zero delay. Points show the pulse-filtered MPS correlators. Solid orange curves show the finite-time Luttinger-liquid fit with free $K_{\mathrm{fit}}$ and $v_{\mathrm{fit}}$. Dashed green curves fix $K$ and $v$ to their Bethe-ansatz values and refit the amplitudes. (e,j) Equal-time envelope of the oscillatory component of $C^{zz}(r,0)$ (points), compared with $(r/r_0)^{-2K_{\mathrm{fit}}}$ (solid orange line) using $K_{\mathrm{fit}}$ from the cross-dephasing fit. Both are normalized at the first plotted separation $r_0$. The probe-inferred power law follows the MPS envelope closely at zero field and decays more steeply at finite field.}
\label{fig:Covariance}
\end{figure*}
Panels (c) and (h) show the downward-only prediction [Eq.~\eqref{eq:Bell_phase_difference_exact}] at $JT_{\max}=40$. At zero field, the sign changes occur closer together as the probe separation increases,  consistent with the characteristic spacing $\pi v/d$ in Eq.~\eqref{eq:Bell_group_delay}. The secular approximation is least controlled at the low-frequency end of the scan. In particular, when $\omega_q T_{\max}\lesssim1$, the coarse-graining window in Eq.~\eqref{eq:Bell_secular_window} cannot be satisfied. The curves in this regime therefore show the downward-only prediction.

We use single-site coupling to examine the spatial decay of the individual spin sectors [panels (e) and (j)]. At $JT_0=3$ and $\omega_q/J=0.25$, we retain separations $d\geq3v_{\mathrm{BA}}T_0$, where the kernel approximately inherits the equal-time spatial dependence [Eq.~\eqref{eq:Bell_pre_lightcone}]. The markers show kernel magnitudes computed from the MPS correlators. At $m=0$, a staggering filter suppresses the smooth longitudinal term and isolates the longitudinal $2k_F$ contribution. The dashed lines use the fixed Bethe--ansatz powers of Eq.~\eqref{eq:Bell_precone_exponents}. At $m=0$, these are $d^{-2/3}$ for the transverse sector and $d^{-3/2}$ for the longitudinal $2k_F$ sector. At $m\simeq0.202$, the transverse reference is $d^{-0.61}$. Only the amplitudes of these reference lines are adjusted to the numerical points.

A separate finite-time fit at $m\simeq0.202$ estimates the scaling dimensions directly. At $\omega_q/J=0.15$, we apply the same finite-time integral [Eq.~\eqref{eq:WAB_finite_time}] to the Luttinger-liquid correlators and fit the resulting kernels over $2\leq d/a\leq40$. With $v=v_{\mathrm{BA}}$, we obtain $x_\parallel\simeq0.80$ and $x_\perp\simeq0.32$, compared with the Bethe--ansatz values $K=0.819$ and $1/(4K)=0.305$.

\subsection{Covariance sensing in the XXZ chain}
\label{sec:XY8_XXZ}

For the dipolar coupling, the longitudinal matter operator appearing in the
generic covariance protocol is
\begin{equation}
B_a^z =\sum_j\left(D_{a,j}^{zz}s_j^z+D_{a,j}^{z+}s_j^+ +D_{a,j}^{z-}s_j^-\right).
\label{eq:Bz_spin_decomp}
\end{equation}
Consequently, although the principal harmonic selects the longitudinal toggling-frame channel, the microscopic dipolar coupling contains all spin components. For a $U(1)$-invariant XXZ bath and a translation-invariant chain, the effective cross spectrum in Eq.~\eqref{eq:chiAB_freq}
is
\begin{align}
\overline S_{AB}^{zz,\rm eff}(\omega;d)&= \int_{-\pi/a}^{\pi/a}\frac{dq}{2\pi}\,e^{iqd} \bigg[|D^{zz}(q;h)|^2\overline S^{zz}(q,\omega)\nonumber\\
&+\frac{|D^{zx}(q;h)|^2}{4}\Big(\overline S^{+-}(q,\omega)+\overline S^{-+}(q,\omega)\Big)\bigg].
\label{eq:Szz_eff}
\end{align}
It is important to note that a longitudinal toggling-frame channel therefore does not imply sensitivity only to microscopic longitudinal spin correlations.

\subsubsection{Large-distance asymptotics}
\label{sec:XY8_LL}

We now evaluate Eq.~\eqref{eq:chiAB_factorized} in the LL regime of the XXZ chain at zero temperature. For antiferromagnetic exchange $J>0$ and $-1<\Delta<1$, the chain is gapless below the saturation field, $h_{\rm sat}=J(1+\Delta)$ \cite{PhysRevB.69.064427}. In particular, $\Delta=1/2$ used below gives $h_{\rm sat}=3J/2$. The pulse sequence probes fluctuations near its principal filter frequency $\omega_\star\simeq\pi/\tau$, where $\tau$ is the interpulse spacing. We take $\omega_\star$ within the low-energy LL regime.

We examine how the covariance changes as the probes are moved apart, to connect the measured signal with the spatial decay of correlations in the chain. At separations large compared with the lattice spacing, $d\gg a$, these correlations follow universal LL scaling laws governed by the interaction-dependent parameter $K$. Two oscillating contributions are relevant in this regime. The longitudinal correlations oscillate at $2k_F=\pi(1-2m)/a$, where $m=\langle s_j^z\rangle$, while the leading transverse correlations are staggered, with wave vector $\pi/a$. We label each contribution by its wave vector $Q_x$ and scaling dimension $x$,
\begin{equation}
(Q_x,x)=\begin{cases}(2k_F,K), & \text{longitudinal},\\[1mm](\pi/a,1/(4K)), & \text{staggered transverse}.
\end{cases}
\label{eq:LL_sectors}
\end{equation}
Here $K=K(\Delta,m)$ sets the correlation exponents and $v=v(\Delta,m)$ is the propagation velocity. The corresponding equal-time envelopes decay as $r^{-2x}$.

To find the spatial decay at the frequency selected by the pulses, we Fourier transform each symmetrized spin correlation in time. The resulting spectrum $\overline S_x(\omega_\star;d)$ has two limits, depending on how the propagation time $d/v$ compares with $1/\omega_\star$. For $a\ll d\ll v/\omega_\star$ we find (see Appendix~\ref{app:LL})
\begin{equation}
\overline S_x(\omega_\star;d)\propto A_x\left(\frac{\omega_\star a}{v}\right)^{2x-1}
\cos(Q_xd+\varphi_x),
\label{eq:Szz_short}
\end{equation}
where $A_x$ and $\varphi_x$ are the amplitude and phase of the spatial oscillation. Only the lattice oscillation depends on $d$ to leading order. For $d\gg v/\omega_\star$, the same expression gives
\begin{align}
\overline S_x(\omega_\star;d)\simeq{} \mathcal F_x(\omega_\star)\left(\frac{a}{d}\right)^x
\cos(Q_xd+\varphi_x)\cos\left(\frac{\omega_\star d}{v}-\frac{\pi x}{2}\right),
\label{eq:Szz_2kF}
\end{align}
with $\mathcal F_x(\omega_\star)\propto
A_x(\omega_\star a/v)^{x-1}$. The first cosine follows the spin pattern along the chain. The second oscillates with the propagation time $d/v$. The envelope decays as $d^{-x}$, with half the equal-time exponent.

These envelopes enter the covariance when the pulse train resolves the oscillations in frequency. For many pulses, $T\gg d/v$, and $|t_d|\ll T$, the cross-dephasing exponent becomes
\begin{equation}
\chi_{AB}(T,d,t_d)\simeq\mathcal W_{\XY}(T)\cos(\omega_\star t_d)\, \overline S_{AB}^{zz,\mathrm{eff}}(\omega_\star;d).
\label{eq:chiAB_asymptotic}
\end{equation}
Here $\mathcal W_{\XY}(T)$ is the weight of the two principal filter peaks at $\pm\omega_\star$, as defined in Eq.~\eqref{eq:app_XY8_weight}.

The effective spectrum includes the smooth $q\simeq0$ contribution as well as the two oscillating contributions. The dipolar weights are $|D^{zz}(q;h)|^2$ for the longitudinal spectrum and $|D^{zx}(q;h)|^2/4$ for the sum of the two transverse spectra. Increasing the probe height $h$ suppresses momenta with $|q|h\gg1$. At positive magnetization, $2k_F<\pi/a$, so the longitudinal contribution is less strongly filtered. It avoids strong geometric suppression when
\begin{equation}
2k_Fh\lesssim 1.
\label{eq:kF_condition}
\end{equation}
For probes directly above lattice sites, the transverse dipolar coupling $D^{zx}$ is odd about each probe, so its lattice Fourier transform vanishes at $\pi/a$. The transverse signal at finite frequency therefore comes from nearby momenta, whose dipolar weights must be kept inside the momentum integral.

When $|\chi_{AB}|\ll1$, Eq.~\eqref{eq:CovarianceXY8} gives
$\mathrm{Cov}_{\XY}\simeq
\tfrac12 e^{-[\chi_{AA}+\chi_{BB}]}\chi_{AB}$. At fixed pulse settings the local attenuation is independent of $d$. If one oscillating contribution can be isolated, its large-distance envelope therefore gives access to $K$ through $d^{-K}$ for longitudinal correlations or $d^{-1/(4K)}$ for staggered transverse correlations. The velocity can be obtained at fixed $d$ by varying the pulse spacing. In Eq.~\eqref{eq:Szz_2kF}, consecutive zeros of the frequency oscillation
are separated by
\begin{equation}
\Delta\omega_\star=\frac{\pi v}{d}.
\label{eq:XY8_velocity_period}
\end{equation}
This spacing measures the propagation time $d/v$. In the parity-symmetric geometry of Sec.~\ref{sec:XY8}, varying the relative pulse delay $t_d$ simply modulates the signal by $\cos(\omega_\star t_d)$. The frequency scan thus gives access to propagation in the bulk, complementing single-probe measurements of boundary-induced recoherence \cite{PhysRevB.111.L161115,}.

\subsubsection{Finite-time covariance and single-site scaling}
\label{sec:num_covariance}

Figure~\ref{fig:Covariance} compares the covariance of dipolar-coupled probes with a single-site benchmark that isolates the longitudinal spin correlations. The top and bottom rows show the XXZ chain at
$\Delta=1/2$ for $h_z/J=0$ and $0.96$, respectively. We use CPMG pulse sequences, which have the same longitudinal switching function as XY8 for matched pulse timings. In both geometries, we evaluate the finite-time integral in Eq.~\eqref{eq:chi_ab} with the full pulse windows.

For dipolar coupling at probe height $h=2a$, we combine the matrix-product-state (MPS) correlators with the dipolar tensor \eqref{eq:dipolar_tensor} and convert the resulting dephasing exponents to the covariance using Eq.~\eqref{eq:CovarianceXY8}. Figures~\ref{fig:Covariance} (a,b,f,g) show the separation and delay dependence for four pulses at $\tau J=6.2$. The longitudinal delay cuts are even in $t_d$, as required by Eq.~\eqref{eq:cov_even}. Their extrema and sign changes vary with separation through the spectrum sampled by Eq.~\eqref{eq:chiAB_even}. The open circles show the full TCL2 calculation for rotating-frame $\pi_x$ pulses with $\omega_q/J=0.5$, with each probe's control phase referenced to the start of its sensing window. In Fig.~\ref{fig:Covariance}(g), the maximum absolute difference from the longitudinal curves is $1.09\times10^{-3}$ across the 36 sampled points. The RMS difference is $6.0\%$ of the RMS longitudinal covariance across all points and $20.2\%$ at $d/a=8$. The full result can differ at opposite delays.

The zero-delay scans in Figs.~\ref{fig:Covariance} (c,h) vary $\omega_\star=\pi/\tau$ at fixed pulse number $N=6$. The frequencies at which the covariance changes sign depend on $d$, as expected from the propagation factor in Eq.~\eqref{eq:Szz_2kF}. In the narrowband limit with one resolved sector, Eq.~\eqref{eq:XY8_velocity_period} relates consecutive zeros to the velocity. Here the finite filter bandwidth and the superposition of spin contributions in Eq.~\eqref{eq:Szz_eff} limit a quantitative application of that relation.

To test whether the probe signal can recover a spatial decay exponent in a geometry that isolates $C^{zz}$, we use the single-site coupling $gS_a^z s_{j_a}^z$. We apply Eq.~\eqref{eq:chi_cross} to both the connected MPS correlator $C^{zz}(d,t)$ and a Luttinger-liquid model containing the oscillatory $2k_F$ term and the smooth longitudinal contribution. We fit the amplitudes, $K$, and the velocity $v$ jointly across sensing times and separations $2\le d/a\le40$. The dashed curves fix $K$ and $v$ to their Bethe-ansatz values at the corresponding magnetization (Appendix~\ref{app:bethe}) and refit the amplitudes.

The longitudinal scaling dimension in Eq.~\eqref{eq:LL_sectors} predicts an equal-time envelope proportional to $r^{-2K}$. We therefore compare the power law set by $K_{\mathrm{fit}}$ with the envelope obtained directly from $C^{zz}(r,0)$ in Figs.~\ref{fig:Covariance} (e,j). At zero field the cross-dephasing fit gives $K_{\mathrm{fit}}=0.745$ and $v_{\mathrm{fit}}/(Ja)=1.238$, close to the Bethe-ansatz values $0.750$ and $1.299$. The inferred exponent $2K_{\mathrm{fit}}=1.489$ is also close to the direct MPS estimate $1.445$. At finite magnetization, the power law inferred from the cross-dephasing fit decays faster than the equal-time MPS envelope, as shown in Fig.~\ref{fig:Covariance}(j).

\section{Implementations}
\label{sec:exp}

The results of Secs.~\ref{sec:XY8_LL} and \ref{sec:Bell_LL} identify two requirements for resolving the algebraic structure of the correlations: the probe coupling must transmit the relevant momentum sector, and the sensing window must reach the finite-time regime, $T\lesssim d/v$ for the light cone and $T\gg d/v+2\pi/\omega$ for the frequency-resolved envelopes. At finite chain temperature $T_{\rm th}$, the zero-temperature algebraic envelopes apply at separations shorter than the thermal
length, $d\ll\ell_{\rm th}=v/(\pi T_{\rm th})$. Frequency-resolved measurements should also lie in the quantum regime, $\omega\gg\pi T_{\rm th}$, where $\omega=\omega_q$ for Bell relaxometry and $\omega=\omega_\star$ for covariance sensing, in units $\hbar=k_B=1$. We now discuss possible platforms.

\emph{Impurity qubits in quantum simulators.} The natural setting replaces the stray-field coupling by an engineered impurity--bath interaction, $V_{\rm imp}=\sum_{a,j,\alpha\beta}g^{\alpha\beta}_{aj}S^\alpha_a s^\beta_j$, whose spatial profile and tensor structure are set by state-dependent collisions, dipolar or microwave-dressed interactions, or Rydberg-mediated couplings. The TCL construction applies whenever the impurity--bath coupling is weak, while the interactions within the simulated system need not be perturbative. Ramsey interferometry of a single impurity immersed in an ultracold bath is established \cite{Adam2022}, multiple impurities have been proposed as probes of nonlocal two-point correlations \cite{Streif2016}, a tunable spin-$1/2$ XXZ Heisenberg model has been realized with ultracold $^7$Li in optical lattices \cite{Jepsen2020}, and programmable XXZ Hamiltonians with tunable anisotropy have been engineered in Rydberg tweezer arrays \cite{Scholl2022}. Three features make these platforms the domain of the present theory. First, the probes couple on the scale of one lattice spacing (the single-site coupling of Sec.~\ref{sec:spinchain}), so there is no $e^{-2|q|h}$ suppression and the staggered $q\simeq\pi/a$ sector, which carries the exponents $K$ and $1/4K$, is accessible alongside the smooth and $2k_F$ sectors at zero field. Second, with $J/\hbar$ in the kHz range, the propagation time $d/v$ is of order milliseconds and is resolved by the control, so that the pre-light-cone regime, the ridge $T\simeq d/v$, and the crossover from the equal-time exponent $2x$ to the frequency-resolved exponent $x$ are all within reach of a single sensing-time scan. Third, $L$, the coupling, $T$, and the qubit or filter frequency can be varied independently, and the system is prepared near its ground state, so the quantum window $\omega\gg\pi T_{\rm th}$ of Sec.~\ref{sec:XY8_LL} is the generic situation rather than an exception. The benchmark of Sec.~\ref{sec:num_covariance} is a direct simulation of this setting and shows what such an experiment returns with $40/J$ of sensing time. For thermodynamic \LL{} behavior one should work in the window $a\ll d\ll L$ with $T\ll t_{\rm rec}$ so that boundary reflections have not returned, and, for the covariance protocol, with enough pulses that $\omega_\star$ lies below the spinon band maximum $\sim v/a$, since otherwise the transverse toggling channels neglected in the principal-harmonic treatment dominate. Spin-1 simulators based on dysprosium or three-state Rydberg systems \cite{Brechtelsbauer2025,Moegerle2025} extend the sensing formalism directly, although the spin-$1/2$ asymptotics derived here must be recomputed---in a gapless spin-1 regime the operator content and scaling dimensions differ, and in the Haldane phase the two-probe signal would instead encode a finite bulk correlation length and the edge modes.

\emph{Solid State Defects.} Solid-state defect spins above a quasi-one-dimensional magnet realize the dipolar geometry of Fig.~\ref{fig:schematic} with
$h\gg a$. The required ingredients exist: NV centers operated as effective spin-$1/2$ sensors with optical readout and microwave control
\cite{RevModPhys.92.015004,Hall2016}, simultaneous XY8 sensing and covariance readout of two spatially resolved NV centers \cite{Rovny2022}, wideband covariance magnetometry \cite{Le2025} and parallel readout of large NV arrays \cite{Cambria2025,Cheng2025}, individually
addressable NV pairs at separations of $10$--$25$~nm
\cite{Dolde2013,Dolde2014,Lee2023,Bersin2019,Hwang2022,Pezzagna2010} and stand-offs down to $\sim8$~nm \cite{Xu2025}; spin-$1/2$ chain compounds with exchange in the $0.2$~meV range and field-tunable Luttinger-liquid regimes are available \cite{Wu2019YbAlO3,Kenzelmann2002Cs2CoCl4,Faure2019BaCo2V2O8}, while the cleanest Heisenberg chains \cite{Motoyama1996} have their weight at $q=\pi/a$ and a large exchange scale. At the distances considered here, the dipolar form factor favors long wavelengths, $|q|\lesssim 1/h$. For $d\simeq20$~nm and $v\simeq200$~m\,s$^{-1}$, the propagation time is $d/v\simeq0.1$~ns, making the front difficult to resolve directly in time. If a single ballistic mode dominates the Bell contrast and the chain parameters stay fixed during the sweep, successive sign changes are separated by $\Delta\omega_q/(2\pi)\simeq v/(2d)\simeq5\,\mathrm{GHz}$ [Eq.~\eqref{eq:Bell_group_delay}], so extracting $d/v$ from their spacing would require tuning the qubit transition over several gigahertz. Whether these oscillations are resolved at the operating temperature depends on the finite-temperature spin dynamics of the compound. In this case, the Bell coherence must in addition be protected against local dephasing, Eq.~\eqref{eq:Bell_linear_dephasing}, since $T_2^*\ll T_1$, and because covariance signals are more strongly affected by readout noise than single-probe observables, enhanced readout such as spin-to-charge conversion is valuable \cite{Rovny2022}.

\section{Conclusions}
\label{sec:conclusions}

We developed a finite-time description of two qubits weakly coupled to an interacting quantum environment. To second order in the probe coupling, the dynamics separates into local decoherence and cross-kernels that encode correlations between the two sensing regions. Bell-state relaxometry isolates the signed cross-relaxation kernel by comparing the joint ground-state populations for two Bell preparations with opposite relative signs. The resulting contrast probes a filtered Wightman correlator, while dynamical-decoupling covariance probes the complementary symmetrized correlator. When probe exchange is a symmetry of the dynamics and the readout, the Bell contrast is even in the preparation phase and its $\pi/2$ quadrature vanishes. Separately, the longitudinal covariance is even in the relative control delay for reflection-equivalent probes with matched pulse trains and shifted sensing windows.

For the gapless XXZ chain, Luttinger-liquid theory predicts two spatial scaling regimes for the oscillating spin sectors of the Bell kernel. Well before the front, $T\ll d/v$, the kernel inherits the equal-time envelope $d^{-2x}$. At sufficiently long sensing times after the front, the kernel resolves the spectrum near the qubit frequency. Within the low-energy regime and at large $\omega_q d/v$, the corresponding envelope decays as $d^{-x}$. The propagation time $d/v$ also sets the oscillation scale of the frequency-resolved signal.

The matrix-product-state calculations benchmark the short-time scaling and the ballistic oscillations. At zero field, the spatial oscillations of the dipolar Bell cross-relaxation rate give $v_{\mathrm{fit}}/(Ja)\simeq1.30$, compared with the Bethe--ansatz value $1.299$. The single-site covariance fit gives $K_{\mathrm{fit}}\simeq0.745$, compared with $K=0.750$. Before the front, the single-site Bell kernels are consistent with the Bethe--ansatz power laws for the staggered transverse sector at both magnetizations and for the longitudinal $2k_F$ sector at zero field. At finite magnetization, a separate fit to the finite-time Bell kernels with $v$ fixed to its Bethe--ansatz value gives $x_\parallel\simeq0.80$ and $x_\perp\simeq0.32$, compared with $0.819$ and $0.305$, respectively.

Single-site coupling can isolate individual spin components and expose their spatial scaling. For dipolar probes, increasing the probe height suppresses large momenta and can favor the smooth contribution, whose oscillations give access to the propagation velocity. The coupling geometry thus determines which signatures of the same many-body correlations are accessible to lattice-scale impurity qubits and spin sensors.

\begin{acknowledgments}
C.P. is an \'{E}cole Normale Sup\'{e}rieure (ENS)-Mitsubishi Heavy Industries (MHI) Chair of Quantum Information supported by MHI. 
\end{acknowledgments}

\appendix

\section{Single-qubit Relaxometry}
\label{app:single}

This appendix collects the single-probe theory of which Sec.~\ref{sec:model} is the two-probe generalization. We consider one spin-$1/2$ probe coupled to the quantum matter through the general tensorial interaction \eqref{eq:Va_def} and subject to an arbitrary control Hamiltonian, and derive its reduced density matrix $\rho_Q(t)$ to second order in the probe--matter coupling without a Markov approximation. The appendix gives the control-dressed kernel and the filter-function form of the dephasing exponent, the exact factorization of the control propagator into free and pulsed segments, the finite-time relaxation kernel, and the Fermi-golden-rule limit in which the time-independent rates of conventional relaxometry are recovered.

\subsection{Reduction to one probe}

The single-probe theory follows from Sec.~\ref{sec:model} by retaining only one probe in Eqs.~\eqref{eq:KAB_def} and \eqref{eq:TCL2_gen_AB}. Its interaction-picture state is $(\widetilde\rho_Q(t)$. Dropping the probe index, the control rotation and dressed kernel are
\begin{align}
\widetilde S^\mu(t)&=\sum_\nu R_{\mu\nu}(t)S^\nu,
\label{eq:R_single}\\
\mathcal K^{\mu\nu}(t_1,t_2)
&=\sum_{\mu',\nu'}R_{\mu'\mu}(t_1)R_{\nu'\nu}(t_2)
C^{\mu'\nu'}(t_1,t_2).
\label{eq:K_single}
\end{align}
For the dipolar coupling to the XXZ chain, $C^{\mu'\nu'}$ is the single-probe version of Eq.~\eqref{eq:Cab_micro}. The static mean field is included in $H_Q$, as in Eq.~\eqref{eq:O_connected}.

For purely longitudinal coupling, $H_{Q,0}=\omega_q S^z$ and ideal transverse $\pi$ pulses give $\widetilde S^z(t)=y_z(t)S^z$. The populations are constant in the interaction picture. The response term in Eq.~\eqref{eq:TCL2_noise_response} vanishes because $[S^z,\{S^z,\rho\}]=[(S^z)^2,\rho]=0$. For stationary noise the coherence obeys $\widetilde\rho_{eg}(T)=\rho_{eg}(0)e^{-\chi_{QQ}(T)}$ within TCL2, with
\begin{align}
\chi_{QQ}(T)
&=\frac12\int_0^T dt_1\int_0^T dt_2\,
y_z(t_1)y_z(t_2)N^{zz}(t_1-t_2)
\nonumber\\
&=\frac12\int_{-\infty}^{\infty}\frac{d\omega}{2\pi}\,
\overline S^{zz}(\omega)|F_z(\omega)|^2 .
\label{eq:filter_function}
\end{align}
Here $\overline S^{zz}(\omega)=\int d\tau\,e^{-i\omega\tau}N^{zz}(\tau)$ and $F_z$ is defined in Sec.~\ref{sec:XY8}. For free precession,
\(|F_z(\omega)|^2=4\sin^2(\omega T/2)/\omega^2\), and
\(\chi_{QQ}(T)/T\to\overline S^{zz}(0)/2\) when the spectrum is continuous
near zero frequency. The finite-time expression can still give nonexponential coherence decay when this asymptotic rate vanishes, as in the local-probe XX-chain example of Ref.~\cite{PhysRevB.111.L161115}.

\subsection{Pulse-driven interaction picture}
\label{app:pulse}

For each probe, the full propagator is the chronological product of the
free and driven segments. For pulses supported on windows
\(I_n=[t_n^-,t_n^+]\), let \(U_f\) denote evolution under \(H_{Q,0}\).
The exact evolution through pulse \(n\) is
\begin{align}
U_{p,n}
&=U_f(t_n^+,t_n^-)\,
\mathcal T\exp\!\left[-i\int_{t_n^-}^{t_n^+}ds\,
H_{c,I}^{(n)}(s)\right],
\label{eq:exact_factorization}\\
H_{c,I}^{(n)}(s)
&=U_f^\dagger(s,t_n^-)\,H_c(s)\,U_f(s,t_n^-).
\nonumber
\end{align}
Composing the segments gives the control rotation in
Eq.~\eqref{eq:Ra_def} in the same product order,
\(R(U_2U_1)=R(U_2)R(U_1)\). In the instantaneous-pulse limit,
\(U_{p,n}\) reduces to the pulse unitary. If the pulse axes are specified in
the free-precession frame, write \(U_Q(t,0)=U_f(t,0)U_c(t)\), with
\(U_c(t)=P_{m(t)}\cdots P_1\) between pulses. The full rotation is then
\(R[U_f(t,0)]R[U_c(t)]\). Appendix~\ref{app:toggling} gives
\(R[U_c(t)]\) for the \(\pi_x\) and \(\pi_y\) pulses used here.

\subsection{Finite-time relaxation}
\label{app:finite_relax}

For a freely precessing probe, \(S^\pm=S^x\pm iS^y\) evolve as
\(e^{\pm i\omega_q t}S^\pm\). Starting in \(|e\rangle\), the leading
transition probability is \(P_{e\to g}(T)=W_\downarrow(T)+O(V^3)\), where
\(W_\downarrow\) is the diagonal kernel of Eq.~\eqref{eq:WAB_finite_time}.
With the later-right convention
\(C^\downarrow(t)=
\langle\delta\mathcal O^-(0)\delta\widetilde{\mathcal O}^+(t)\rangle_M\),
define \(S^\downarrow(\omega)=\int dt\,e^{-i\omega t}C^\downarrow(t)\).
Stationarity gives
\begin{align}
W_\downarrow(T)
&=\frac14\int_{-T}^{T}dt\,(T-|t|)
e^{-i\omega_qt}C^\downarrow(t)
\nonumber\\
&=\frac14\int_{-\infty}^\infty\frac{d\omega}{2\pi}\,
S^\downarrow(\omega)
\left|\int_0^T dt\,e^{i(\omega-\omega_q)t}\right|^2 .
\label{eq:W_single}
\end{align}
The sensing window thus resolves a band of width of order \(1/T\) about
\(\omega_q\).

\subsection{Fermi-golden-rule limit}
\label{app:fgr}

Differentiating Eq.~\eqref{eq:W_single} gives the instantaneous
second-order coefficient
\begin{equation}
\Gamma_\downarrow(T)=\frac12\Re\int_0^T dt\,e^{-i\omega_qt}
C^\downarrow(t).
\label{eq:Gamma_finite}
\end{equation}
If this integral converges, its long-time limit is
\(\Gamma_\downarrow^{\rm FGR}=S^\downarrow(\omega_q)/4\)
\cite{PhysRevB.98.195433,PhysRevB.99.104425,PhysRevLett.121.187204}.
The excitation coefficient is
\(\Gamma_\uparrow^{\rm FGR}=S^\uparrow(-\omega_q)/4\), where
\(S^\uparrow\) is defined by interchanging \(+\) and \(-\) in
\(C^\downarrow\). In the constant-rate population equation
\(\dot P_e=-\Gamma_\downarrow^{\rm FGR}P_e+
\Gamma_\uparrow^{\rm FGR}(1-P_e)\), the population approaches its
stationary value exponentially with
\(T_1^{-1}=\Gamma_\downarrow^{\rm FGR}+\Gamma_\uparrow^{\rm FGR}\).
For a bath with a finite memory time \(\tau_B\), the linear transition
probability applies in the window \(\tau_B\ll T\ll T_1\).
The constant-rate equation on times of order \(T_1\) uses a coarse-graining
time between these two scales. In a finite bath, the spectrum must also
be effectively continuous at the resolution \(1/T\).

\section{Second-order reduced map}
\label{app:dyson}

Expanding the interaction-picture evolution of the factorized initial state
in Eq.~\eqref{eq:init} gives the second-order contribution
\begin{align}
\Phi_2^{AB}(T)[\rho]
&=-\int_0^T dt_1\int_0^{t_1}dt_2
\nonumber\\
&\quad\times\Tr_M\!\left[
\widetilde V(t_1),
[\widetilde V(t_2),\rho\otimes\rho_M]
\right].
\label{eq:Phi2_def}
\end{align}
The first-order term vanishes because the matter operators are centered.
To evaluate the trace, write
\(\widetilde V(t)=\sum_{a,\mu}S_a^\mu\otimes\delta\mathcal O_{a,\rm d}^\mu(t)\), with
\(\delta\mathcal O_{a,\rm d}^\mu(t)=\sum_\lambda R_{\lambda\mu}^{(a)}(t)
\delta\widetilde{\mathcal O}_a^\lambda(t)\) denoting the control-dressed operator.
Cyclicity of the matter trace gives the two orderings
\(\mathcal K_{ab}^{\mu\nu}(t_1,t_2)\) and
\(\mathcal K_{ba}^{\nu\mu}(t_2,t_1)=
[\mathcal K_{ab}^{\mu\nu}(t_1,t_2)]^*\).
They multiply \([S_a^\mu,S_b^\nu\rho]\) and
\([\rho S_b^\nu,S_a^\mu]\), respectively. Differentiating the map with
respect to \(T\) gives Eq.~\eqref{eq:TCL2_gen_AB}.
At fixed \(T\), the first difference between the time-ordered TCL2
propagator and \(e^{\Phi_2^{AB}(T)}\) is of fourth order in \(V\).
The two propagators coincide within TCL2 whenever the generators at
different times commute.

\section{Toggling-frame control matrix for ideal \texorpdfstring{\(\pi\)}{pi} pulses}
\label{app:toggling}

For \(P_\nu=e^{-i\pi S^\nu}\) with \(\nu=x,y\), the control rotations are
\(R_x(\pi)=\diag(1,-1,-1)\) and \(R_y(\pi)=\diag(-1,1,-1)\).
If \(m_x(t)\) and \(m_y(t)\) count the pulses applied about each axis before
time \(t\), their product gives
\begin{align}
R_{\rm tog}(t)&=\diag\big(y_x(t),y_y(t),y_z(t)\big),
\nonumber\\
y_x(t)&=(-1)^{m_y(t)},\qquad y_y(t)=(-1)^{m_x(t)},
\nonumber\\
y_z(t)&=(-1)^{m_x(t)+m_y(t)}.
\label{eq:app_toggling_signs}
\end{align}
In particular, every pulse flips \(y_z\). The signs apply independently
to each probe and describe the pulse-only part of its control matrix.

\section{Filter functions of the pulse trains}
\label{app:filters}

For \(N\) pulses at \(u_r=(r-\tfrac12)\tau\), \(1\le r\le N\), set
\(u_0=0\) and \(u_{N+1}=T=N\tau\). Integrating the signs in
Eq.~\eqref{eq:app_toggling_signs} over the free intervals gives
\begin{equation}
F_z(\omega)=\frac{1}{i\omega}\sum_{r=0}^{N}(-1)^r
\left(e^{i\omega u_{r+1}}-e^{i\omega u_r}\right).
\label{eq:Fz_generic}
\end{equation}
For CPMG with \(\pi_x\) pulses, \(F_y=F_z\) and
\(F_x=(e^{i\omega T}-1)/(i\omega)\). For one XY8 cycle
\((x,y,x,y,y,x,y,x)\), the signs on the nine free intervals are
\begin{align}
q_x&=(+,+,-,-,+,-,-,+,+),\nonumber\\
q_y&=(+,-,-,+,+,+,-,-,+),\nonumber\\
q_z&=(+,-,+,-,+,-,+,-,+).
\nonumber
\end{align}
For \(M\) cycles, the component filters are
\begin{align}
F_\mu(\omega)
&=f_\mu^{\XY}(\omega;\tau)\sum_{m=0}^{M-1}e^{i8m\omega\tau},
\nonumber\\
f_\mu^{\XY}(\omega;\tau)
&=\frac{1}{i\omega}\sum_{r=0}^{8}q_{\mu,r}
\left(e^{i\omega u_{r+1}}-e^{i\omega u_r}\right),
\end{align}
where \(u_9=8\tau\) in the single-cycle expression. The longitudinal
filter is Eq.~\eqref{eq:Fz_generic} with \(N=8M\). The pulse-only
transverse filters vanish at \(\omega_\star=\pi/\tau\) and the width of
the principal longitudinal peak scales as \(1/T\). When that peak and
its negative-frequency partner dominate Eq.~\eqref{eq:chiAB_even},
the weight \(\mathcal W_{\XY}\) is
\begin{align}
\mathcal W_{\XY}(T)
&=\frac12\int_{\mathcal B_\star}\frac{d\omega}{2\pi}
|F_z(\omega)|^2 ,
\label{eq:app_XY8_weight}
\end{align}
where \(\mathcal B_\star\) contains both principal peaks. Extending the
integral to all frequencies gives \(T/2\) by Parseval's identity for
ideal pulses. This extension counts the higher harmonics as well. For many ideal
cycles, the fundamental pair alone carries \(8/\pi^2\) of the total
filter power, giving \(\mathcal W_{\XY}(T)\simeq 4T/\pi^2\).

\section{Derivation of the covariance formula}
\label{app:cov}

Under the longitudinal reduction in Eq.~\eqref{eq:Veff_XY8}, let
\(y_a=y_z^{(a)}\), including the relative delay in \(y_B\). For the
product-basis coherence \(\rho_{mn}\), write
\(\Delta s_a=s_a^m-s_a^n\) and \(\Sigma s_a=s_a^m+s_a^n\), where
\(s_a^m=\pm\tfrac12\) is the eigenvalue of \(S_a^z\) in \(|m\rangle\).
Equation~\eqref{eq:TCL2_noise_response} then gives
\begin{align}
\dot\rho_{mn}(t)&=-\Lambda_{mn}(t)\rho_{mn}(t),\nonumber\\
\Lambda_{mn}(t)&=\sum_{a,b}\Delta s_a\int_0^t dt'\,
y_a(t)y_b(t')
\nonumber\\
&\quad\times\big[
\Delta s_b N^{zz}_{ab}(t,t')+
i\Sigma s_b X^{zz}_{ab}(t,t')\big].
\label{eq:app_Lambda}
\end{align}
Each coherence therefore acquires the scalar exponent
\(-\int_0^Tdt\,\Lambda_{mn}(t)\).

Use \(|0\rangle\) for the \(+\tfrac12\) eigenstate and \(|1\rangle\)
for the \(-\tfrac12\) eigenstate. The joint readout
\(\langle S_A^y S_B^y\rangle_T=
\tfrac12\Re[\rho_{01,10}(T)-\rho_{00,11}(T)]\)
involves only coherences with \(\Sigma s_A=\Sigma s_B=0\).
The symmetry \(N_{ab}^{zz}(t_1,t_2)=N_{ba}^{zz}(t_2,t_1)\) then
reduces the integrated real exponents to
\(\chi_{AA}+\chi_{BB}\mp2\chi_{AB}\) for
\(\rho_{01,10}\) and \(\rho_{00,11}\), respectively, with \(\chi_{ab}\)
defined in Eq.~\eqref{eq:chi_ab}. Both coherences start at \(1/4\).
In the marginal readout of probe \(A\), the two coherences
\(\rho_{0j,1j}\) have the same attenuation \(e^{-\chi_{AA}}\) and
opposite response phases, since \(s_B^0=-s_B^1\). Their imaginary
parts cancel, and the same holds for probe \(B\). The connected
covariance therefore equals the joint readout and yields
Eq.~\eqref{eq:CovarianceXY8}.

For numerical evaluation, let \(I_r^{(a)}\) be the intervals within
\([0,T]\) on which \(y_a\) is constant, with sign \(y_{a,r}\).
The exact longitudinal filter can then be applied directly in time,
\begin{align}
\chi_{AB}(T,t_d)
&=\frac12\sum_{r,s}y_{A,r}y_{B,s}
\nonumber\\
&\quad\times\int_{I_r^{(A)}}dt_1\int_{I_s^{(B)}}dt_2\,
N_{AB}^{zz}(t_1,t_2).
\label{eq:chi_cross}
\end{align}
The intervals on \(B\) incorporate the delay and the endpoints of the
sensing window.

\section{Derivation of the Bell-state signal}
\label{app:bell}
\label{app:bell_deph}

We retain the longitudinal and probe-frequency-secular relaxation terms
of Eq.~\eqref{eq:TCL2_gen_AB}. The mixed and anomalous terms oscillate
at \(\omega_q\) and \(2\omega_q\). They can be averaged out when there
is a coarse-graining window
\begin{equation}
\omega_q^{-1}\ll\Delta t\ll\min\{T,\tau_{\rm var}\},
\label{eq:Bell_secular_window}
\end{equation}
where \(\tau_{\rm var}\) is the shortest time scale of the retained
dynamics or features to be resolved. The finite-time coefficients of
the retained terms are kept throughout. To neglect upward transitions,
we require
\begin{equation}
\max_{a=A,B}\sup_{0<t\le T} W^\uparrow_{aa}(t)\ll1 .
\label{eq:Bell_upward_condition}
\end{equation}
Here \(W^\uparrow_{ab}\) follows from Eq.~\eqref{eq:Wab_double} by
\(\omega_q\mapsto-\omega_q\) and interchanging the circular components
\(+\) and \(-\). The matrix \(W^\uparrow\) is a Gram matrix, so
\(|W^\uparrow_{AB}|^2\le W^\uparrow_{AA}W^\uparrow_{BB}\).
A finite sensing window can give off-resonant excitation even for a
bath in its ground state.

For a stationary bath we use the later-right correlator
\begin{equation}
C_{ab}^{\downarrow}(t;d_{ab})=
\big\langle\delta\mathcal O_a^-(0)
\delta\widetilde{\mathcal O}_b^+(t)\big\rangle_M .
\label{eq:Cdown_generic}
\end{equation}
For mirror-equivalent probes, probe exchange and Hermitian conjugation
give
\begin{align}
C_{AB}^{\downarrow}(t;d)
&=C_{BA}^{\downarrow}(t;d)
=\big[C_{AB}^{\downarrow}(-t;d)\big]^*,
\nonumber\\
W_{AB}^{\downarrow}(T;d)
&=W_{BA}^{\downarrow}(T;d)\in\mathbb R .
\label{eq:parity_Bell}
\end{align}

In the one-excitation basis \(|A\rangle=|eg\rangle\),
\(|B\rangle=|ge\rangle\), the triangular integral in the downward
generator is
\begin{align}
M_{ab}^{\downarrow}(t)
&=\frac14\int_0^t dt_1\int_0^{t_1}dt_2\,
e^{i\omega_q(t_1-t_2)}
\nonumber\\
&\quad\times
\big\langle\delta\widetilde{\mathcal O}_a^-(t_1)
\delta\widetilde{\mathcal O}_b^+(t_2)\big\rangle_M .
\label{eq:app_Mdown}
\end{align}
The complementary triangle is its Hermitian adjoint, so
\(M^\downarrow+M^{\downarrow\dagger}=W^\downarrow\).
The projected downward generator acts on the one-excitation block
\(\rho_1\) as
\(\dot\rho_1|_\downarrow=-\dot M^\downarrow\rho_1
-\rho_1\dot M^{\downarrow\dagger}\), with the lost population transferred
to \(|gg\rangle\). Exchange symmetry gives
\(M^\downarrow=M_0^\downarrow \mathds{1}+M_1^\downarrow\sigma_x\), where
\(\sigma_x=|A\rangle\langle B|+|B\rangle\langle A|\),
\(2\Re M_0^\downarrow=\overline W\), and
\(2\Re M_1^\downarrow=W_{AB}^\downarrow\).

For the Bell preparation in Eq.~\eqref{eq:bell_state}, define
\(p=\Tr\rho_1\) and \(x_{\rm B}=\Tr(\sigma_x\rho_1)\). These variables
have initial values \(p(0)=1\) and \(x_{\rm B}(0)=\cos\phi\), and
\(P_{gg}^{(\phi)}=1-p\). The coherent exchange from the anti-Hermitian
part of \(M^\downarrow\) commutes with both observables. The longitudinal
term follows from Eq.~\eqref{eq:app_Lambda} with \(y_a=1\).
It leaves \(p\) unchanged and damps the one-excitation coherence with
the exponent \(\chi_R-2\chi_{AB}\), where \(\chi_R=\chi_{AA}+\chi_{BB}\).
The resulting equations are
\begin{align}
\dot p&=-\dot{\overline W}p-\dot W_{AB}^\downarrow x_{\rm B},
\nonumber\\
\dot x_{\rm B}&=-\dot W_{AB}^\downarrow p
-\big(\dot{\overline W}+\dot\chi_R-2\dot\chi_{AB}\big)x_{\rm B}.
\label{eq:app_bell_dynamics}
\end{align}

Without longitudinal dephasing, \((p\pm x_{\rm B})/2\) evolve
independently with accumulated exponents
\(\overline W\pm W_{AB}^\downarrow\). Their initial weights
\((1\pm\cos\phi)/2\) give Eqs.~\eqref{eq:Bell_population_exact}
and \eqref{eq:Bell_phase_difference_exact}. With dephasing, let
\(\Delta p\) and \(\Delta x_{\rm B}\) be the half-differences between
phases \(\phi\) and \(\phi+\pi\). At zeroth order in the cross-kernels,
\(\Delta p=0\) and
\(\Delta x_{\rm B}(t)=\cos\phi\,e^{-\overline W(t)-\chi_R(t)}\).
Using this coherence as the source in the first line of
Eq.~\eqref{eq:app_bell_dynamics} gives the leading cross contribution,
\begin{align}
\Delta P_{gg}^{(\phi)}(T,d)
&\simeq e^{-\overline W(T)}\cos\phi
\int_0^T dt\,e^{-\chi_R(t)}\,\dot W_{AB}^\downarrow(t;d).
\label{eq:duhamel}
\end{align}
The first cross-dephasing correction is proportional to both cross
dephasing and cross relaxation. If \(W_{AB}^\downarrow(t)\)
is proportional to \(\chi_R(t)\) throughout the interval, the integral
reduces to
\(W_{AB}^\downarrow(T)[1-e^{-\chi_R(T)}]/\chi_R(T)\), with its continuous
limit at \(\chi_R(T)=0\). Otherwise, Eq.~\eqref{eq:duhamel} retains the
time dependence of the coherence at the instant of each relaxation
contribution. The distinction from \(e^{\Phi_2^{AB}}\) can matter when
the local exponents are of order unity.

\section{Bethe-ansatz reference values}
\label{app:bethe}

For the XXZ Hamiltonian \eqref{eq:HB} with \(\Delta=\cos\gamma\), define
the kernels and convolution on the occupied rapidity interval by
\begin{align}
a_n(\lambda)&=\frac{\gamma}{2\pi}
\frac{\sin(n\gamma)}{\cosh(\gamma\lambda)-\cos(n\gamma)},\nonumber\\
(a_n*f)(\lambda)&=\int_{-B}^{B}d\lambda'\,
a_n(\lambda-\lambda')f(\lambda').
\end{align}
The root density \(\rho\), dressed energy \(\varepsilon\), and dressed
charge \(Z\) obey \cite{Takahashi1999}
\begin{align}
\rho+a_2*\rho&=a_1,\nonumber\\
\varepsilon+a_2*\varepsilon
&=h_z-2\pi J\frac{\sin\gamma}{\gamma}a_1,\nonumber\\
Z+a_2*Z&=1,
\end{align}
with the Fermi rapidity \(B\) fixed by \(\varepsilon(\pm B)=0\).
The bulk parameters are
\begin{align}
m&=\frac12-\int_{-B}^{B}d\lambda\,\rho(\lambda),
& K&=Z(B)^2,\nonumber\\
v&=\frac{a\,\varepsilon'(B)}{2\pi\rho(B)},
& \varepsilon_F&=-\varepsilon(0).
\end{align}
Here \(a\) is the lattice spacing, and \(2k_F=\pi(1-2m)/a\).
At \(m=0\), \(v=\pi Ja\sin\gamma/(2\gamma)\) and
\(K=\pi/[2(\pi-\gamma)]\), giving \(v/(Ja)=1.299\) and \(K=0.750\)
for \(\Delta=1/2\). At \(m=28.5/141\simeq0.202\), the same anisotropy
gives \(h_z/J=0.970\), \(v/(Ja)=0.975\), \(K=0.819\),
\(2k_Fa=1.872\), and \(\varepsilon_F/J=0.49\).
The dressed band depth \(\varepsilon_F\) sets an upper energy scale
for the low-energy fits. It decreases towards zero as
\(h_z\) approaches \(h_{\rm sat}=J(1+\Delta)\).

\section{Luttinger-liquid scaling of the filtered correlators}
\label{app:LL}

The two oscillating sectors have wave number $Q_x$ and scaling
dimension $x$ as in Eq.~\eqref{eq:LL_sectors}. The coupling projects onto these sectors through the form factors in
Eqs.~\eqref{eq:Szz_eff} and \eqref{eq:Bplus_spin_decomp}.
For the Bell kernel, assume first that the appropriate form factor
$\mathcal D_x(q;h)$ is nearly constant over the momenta sampled
around $Q_x$. Insert the zero-temperature Wightman function
\eqref{eq:LL_Wightman_zeroT} into Eq.~\eqref{eq:WAB_finite_time}.
With \(u=vt/d\), \(\eta=vT/d\), and \(z=\omega_qd/v\), the triangular
window contributes \((d/v)^2(\eta-|u|)\,du\), and the correlator
contributes \((a/d)^{2x}\). This gives Eq.~\eqref{eq:WAB_scaling_form},
with the dimensionless scaling function
\begin{align}
\mathcal F_x(\eta,z)
&=\int_{-\eta}^{\eta}du\,(\eta-|u|)e^{-izu}
\nonumber\\
&\quad\times\big[(1-u-i0)(1+u+i0)\big]^{-x}.
\label{eq:Bell_scaling_function}
\end{align}
The infinitesimal prescription is the $a/d\to0^+$ limit of the
short-distance cutoff in Eq.~\eqref{eq:LL_Wightman_zeroT}.
The integrand at \(-u\) is the complex conjugate of that at \(u\),
so the scaling function is real. For \(\eta\ll1\), replacing the
correlator by its equal-time value gives
\(\mathcal F_x(\eta,z)\simeq4\sin^2(z\eta/2)/z^2\).

In the frequency-resolved regime, the later-right spectrum is supported
at \(\omega>0\). At fixed \(\omega\) the low-energy threshold weight
near \(Q_x\) is proportional to
\((\omega^2-v^2k^2)^{x-1}\) for \(|k|<\omega/v\), with \(k=q-Q_x\).
The spatial transform of one peak with a constant form factor
therefore reduces, with \(s=vk/\omega\), to
\begin{align}
&\int_{-\omega/v}^{\omega/v}dk\,e^{ikd}
(\omega^2-v^2k^2)^{x-1}
\nonumber\\
&\qquad=\frac{\omega^{2x-1}}{v}
\int_{-1}^{1}ds\,e^{izs}(1-s^2)^{x-1},
\qquad z=\frac{\omega d}{v}.
\end{align}
The last integral equals
\(\sqrt{\pi}\,\Gamma(x)(2/z)^{x-1/2}J_{x-1/2}(z)\).
Including the carrier oscillation and the coupling weight gives,
up to a sector-dependent normalization,
\begin{align}
S_{AB,x}^{\downarrow}(d,\omega)
&\propto|\mathcal D_x(Q_x;h)|^2\cos(Q_xd+\varphi_x)
\nonumber\\
&\quad\times\frac{\omega^{2x-1}}{v}
\left(\frac{2}{z}\right)^{x-\frac12}J_{x-\frac12}(z).
\label{eq:Bell_Bessel}
\end{align}
At \(z\ll1\), the Bessel factor tends to a constant, leaving the
\(\omega^{2x-1}\) dependence and the carrier oscillation.
At \(z\gg1\), it gives the envelope \(d^{-x}\) and the oscillation
\(\cos(\omega d/v-\pi x/2)\). These limits also give the corresponding
positive-frequency symmetrized spectra in
Eqs.~\eqref{eq:Szz_short} and \eqref{eq:Szz_2kF}, with the appropriate
sector weights. When a form factor varies appreciably over the sampled
momenta or vanishes at \(Q_x\), its momentum dependence must stay inside
the spatial transform. This applies to the site-centered \(D^{zx}(q;h)\)
weight at \(Q_x=\pi/a\) in the covariance protocol.

\section{Tensor-network benchmark}
\label{app:numerics}
\label{sec:num_setup}

We use open XXZ chains with \(\Delta=1/2\), taking \(L=140\) at \(m=0\)
and \(L=141\) in the sector \(S_{\rm tot}^z=28.5\).
We obtain the ground state by DMRG in the chosen magnetization sector,
apply \(s_{j_0}^{\kappa}\) at a central site \(j_0\), and evolve
the resulting state by fourth-order TEBD with time step \(0.05/J\)
and bond dimension up to \(640\) \cite{tenpy2024}.
This gives the correlators
\begin{align}
C^{\lambda\kappa}(r,t)
&=\big\langle\delta\widetilde s_{j_0+n}^{\lambda}(t)
\delta s_{j_0}^{\kappa}(0)\big\rangle_M,
\nonumber\\[-0.25ex]
r&=na,\qquad |r|\le60a,\quad 0\le t\le40/J .
\label{eq:num_corr}
\end{align}
At \(m=0\), spin-flip symmetry gives
\(C^{+-}=C^{-+}=2C^{xx}\).

The eigenstates within a fixed sector are independent of \(h_z\). For the later-left convention in Eq.~\eqref{eq:num_corr}, the field dependence is
\begin{align}
C^{\lambda\kappa}(r,t;h_z+\delta h_z)
&=e^{-iq_\lambda\,\delta h_z t}
C^{\lambda\kappa}(r,t;h_z),
\label{eq:app_field_rephase}
\end{align}
with \(q_z=0\) and \(q_\pm=\pm1\). For Fig.~\ref{fig:Bell_relaxometry}, we use Eq.~\eqref{eq:app_field_rephase} to rephase the finite-magnetization transverse correlators from $h_z/J=0.96$ to the thermodynamic field $h_z(m)/J\simeq0.970$ from Appendix~\ref{app:bethe}. Figure~\ref{fig:Covariance} uses the correlators at $h_z/J=0.96$. The largest discarded weight per TEBD step is below $10^{-5}$ at finite
magnetization, and the on-site sum rules \(C^{-+}(0,0)=\tfrac12-\langle s^z_{j_0}\rangle_M\) and \(C^{zz}(0,0)=\tfrac14-\langle s^z_{j_0}\rangle_M^2\) are satisfied to $10^{-4}$ in both runs.

To construct the probe kernels, we approximate bulk spin correlators in Eq.~\eqref{eq:Cab_micro} by the centre-source data at the same separation. For the Bell kernel the later-right ordering follows from stationarity, which moves the time argument from the
right operator to the left with a change of sign. Negative times
are reconstructed by conjugation with the operators and sites
interchanged.

The protocol kernels are evaluated using the full time windows in
Eqs.~\eqref{eq:chi_cross} and \eqref{eq:WAB_finite_time}. We use
the dipolar weights in Eq.~\eqref{eq:dipolar_tensor} or single-site
couplings that isolate the desired chain-spin sector. For Bell
relaxometry, \(g\,\mathbf S_a\cdot\mathbf s_{j_a}\) selects the
transverse chain sector in the relaxation channel, while
\(gS_a^x s_{j_a}^z\) selects the longitudinal sector. For the
longitudinal covariance benchmark, the corresponding single-site
coupling is \(gS_a^z s_{j_a}^z\).
As a reference we evaluate \(e^{\Phi_2^{AB}(T)}\) on the
16-dimensional two-qubit Liouville space with all Cartesian
channels and the full control rotations, including free precession.
For Fig.~\ref{fig:Covariance}, the dipolar calculations use
$g_{\rm dip}/(Ja^3)=0.5$ and $h/a=2$, and the single-site calculations
use $g/J=0.5$. The four-pulse finite-field delay scan in
Fig.~\ref{fig:Covariance}(g) gives $\chi_{AA}=0.0895$.

\bibliographystyle{apsrev4-2}
\bibliography{bibliography}

\end{document}